\documentclass[a4paper,11pt]{article}
\pdfoutput=1 

\usepackage{jheppub} 

\usepackage[T1]{fontenc} 
\usepackage{booktabs}
\usepackage{slashed}
\usepackage{nicefrac}
\usepackage{graphicx}
\usepackage{mathtools}
\usepackage{xspace}
\usepackage{float}
\usepackage{mathrsfs}
\usepackage{multirow}
\usepackage{xcolor}

\def\baa{\begin{array}}
\def\eaa{\end{array}}

\usepackage{soul}
\definecolor{hlyellow}{HTML}{FFF58D}
\sethlcolor{hlyellow}
\newcommand{\met}{{\slashed{E}_T}} 
 
\newcommand{\gev}{\textrm{\;GeV}}

\newcommand{\sigg}{\sigma_{>}}
\newcommand{\sigl}{\sigma_{<}}

\newcommand{\epchrg}{\varepsilon_{\rm charge}}
\newcommand{\abinv}{\textrm{ab}^{-1}}
\newcommand{\fbinv}{\textrm{fb}^{-1}}
\newcommand{\ttbar}{{t\bar{t}}}
\newcommand{\bbbarz}{{b\bar{b}Z}}
\newcommand{\stt}{\sigma_{\ttbar}}
\newcommand{\sbbz}{\sigma_{\bbbarz}}
\newcommand{\zbb}{Zb\bar{b}}
\newcommand{\bbll}{b\bar{b}\ell^+\ell^-}

\newcommand{\mll}{m_{\ell\ell}}

\newcommand{\lag}{\mathscr{L}}
\newcommand{\mcM}{\mathcal{M}}
\newcommand{\mcP}{\mathcal{P}}
\newcommand{\swsq}{\sin^2\theta_W}
\newcommand{\thw}{\theta_W}
\newcommand{\sw}{s_W}
\newcommand{\cw}{c_W}
\newcommand{\sm}{\text{sm}}

\newcommand{\propg}{\mcP_{\gamma}}
\newcommand{\propz}{\mcP_{Z}}
\newcommand{\madgraph}{\texttt{MG5\_aMC@NLO}\xspace}
\newcommand{\pythia}{\texttt{Pythia8}\xspace}
\newcommand{\fastjet}{\texttt{fastjet}\xspace}
\newcommand{\hepmc}{\texttt{HepMC}\xspace}
\newcommand{\mcfm}{\texttt{MCFM}\xspace}
\newcommand{\delphes}{\texttt{Delphes}\xspace}

\newcommand{\dif}{{\textsc{df}}}
\newcommand{\saf}{{\textsc{sf}}}

\title{\boldmath Complementary constraints on $Zb\bar{b}$ couplings at the LHC}

\author[b]{Fady Bishara}
\author[a]{and Zhuoni Qian}

\affiliation[a]{School of Physics, Hangzhou Normal University, Hangzhou, Zhejiang 311121, China}
\affiliation[b]{Deutsches Elektronen-Synchrotron DESY, Notkestr. 85, 22607 Hamburg, Germany}

\emailAdd{fady.bishara@desy.de}
\emailAdd{zhuoniqian@hznu.edu.cn}

\graphicspath{{./figs/}}
\preprint{DESY 23-086}

\abstract{
We propose a new strategy to probe the $Z$ boson couplings to bottom and charm quarks at the LHC. In this work we mainly focus on the case of bottom quarks. Here, the $Z$ boson is produced in association with two $b$-jets and decays to electrons or muons. In this final state, tagging the charge of the $b$-jets allows us to measure the charge asymmetry and thus to directly probe the $Zb\bar{b}$ couplings. The leptonic final state not only allows us to cleanly reconstruct the $Z$ boson but also to mitigate the otherwise overwhelming backgrounds. Furthermore, while LEP could only scan a limited range of dilepton invariant masses, there is no such limitation at the LHC. Consequently, this allows us to make full use of the interference between the amplitudes mediated by a $Z$ boson and a photon. Using the full high-luminosity LHC dataset of 3$~\abinv$ and with the current flavor and charge-tagging capabilities would allow us to reject the wrong-sign right-handed coupling solution by 4$\sigma$. Further improving the charge-tagging efficiency would disfavor it by 6$\sigma$.
}

\begin{document} 
\maketitle
\flushbottom

\section{Introduction} 
\label{sec:intro}
The wealth of remarkably precise measurements of the Standard Model (SM)  electroweak parameters is an enduring legacy of LEP~\cite{ALEPH:2001mdb,ALEPH:2005ab,ALEPH:2009ogm,ALEPH:2010aa}.
In this treasure trove, a handful of anomalies have withstood the test of time.
Among them, the bottom-quark forward-backward asymmetry at the $Z$-pole, $A_{FB}^{0,b}$, remains in tension with the SM prediction at $\gtrsim 2\sigma$.\footnote{Given the variation in fitting strategies and choice of inputs, different groups find slightly differing values for the tension with the SM, even when they use the same input value for $A_{FB}^b$ based on the next-to-next-to-leading order in the strong coupling constant with massive $b$-quarks in ref.~\cite{Bernreuther:2016ccf}. For example, the \texttt{Gfitter} group finds a pull of 2.4$\sigma$~\cite{Haller:2018nnx}, a more recent fit by the \texttt{HEPfit} collaboration finds a pull of 2.2$\sigma$~\cite{deBlas:2021wap}, and the 2022 edition of the PDG finds $2\sigma$~\cite{ParticleDataGroup:2022pth}.}
Since the forward backward asymmetry depends on the chiral nature of the $Z$ boson couplings, it presents an interesting challenge from a model building standpoint.
This challenge is further complicated by the fact that the ratio of the partial decay width into $b$-quarks to the total hadronic width of the $Z$, $R_b^0$, is in good agreement with the SM.
Thus, satisfying both observables imposes non-trivial constraints on the left- and right-handed $Zb\bar{b}$ couplings; namely, that both should receive contributions from new physics beyond the SM~\cite{Gori:2015nqa}.
Nevertheless, this anomaly has inspired many new-physics models including ones with vector-like bottom quarks~\cite{Choudhury:2001hs,He:2003qv, Cheung:2020vqm,Crivellin:2020oup}, additional gauge bosons~\cite{Agashe:2006at,Liu:2017xmc}, and composite Higgs models \cite{DaRold:2010as,Alvarez:2010js,Andres:2015oqa}.

Future lepton colliders in general, and the FCC-ee in particular, will produce orders of magnitude more $Z$ bosons than LEP and will consequently probe the $Zb\bar{b}$ couplings with exquisite precision~\cite{Gori:2015nqa}.
However, while progress is being made, none of these machines have yet to be approved.
On the other hand, an electron ion collider (EIC) at Brookhaven was approved and projections for its capability and that of other EIC proposals to probe these couplings was investigated in~\cite{Yan:2021htf, Li:2021uww}.
In the case of hadron colliders, neither the Tevatron \cite{CDF:2015rth,CDF:2016wlu} nor the large hadron collider (LHC)~\cite{LHCb:2014jms}
have so far been able to provide competitive bounds, mostly due to large QCD backgrounds and large theory uncertainties; see also~\cite{Murphy:2015cha}.

Still, there are many proposals in the literature to probe the $Zb\bar{b}$ couplings at the LHC.
For example, the $Z$ polarization asymmetry in the associated production of a $Z$ boson with a $b$-quark was studied in~\cite{Beccaria:2012xw}.
Leveraging the excellent charge tagging capabilities at LHCb, the authors of~\cite{Gauld:2015qha,Gauld:2019doc} constructed a ratio of the bottom to charm asymmetry in the difference of rapidities of the heavy quark and its anti-quark.
Indirect sensitivity can also be gained via coupling to a $b$-quark in the loop-induced $gg\to Zh$ process in ref.~\cite{Yan:2021veo} which claims a comparable bound to LEP at the high luminosity LHC (HL-LHC).
Another interesting proposal~\cite{Breso-Pla:2021qoe} studies the Drell-Yan forward backward asymmetry of the light quarks at LHC. There, the authors exploit the dependence of the differential cross-section on the di-lepton rapidity to disentangle the left- and right-handed up and down quark couplings.

One important point worth mentioning here is that there is a sign ambiguity of the couplings, $g_{b,L/R} \to -g_{b,L/R}$.
While the wrong-sign solution for the left-handed coupling, $g_L$, is strongly disfavored by the limited off-shell $A^{0,b}_{FB}$ data from LEP~\cite{Choudhury:2001hs}, the $g_R$ ambiguity remains.
Resolving it clearly requires linear sensitivity to the coupling.
This arises from the interference between the $Z$ and photon-mediated amplitudes which must be of the same order to maximize the interference term. This is indeed the case in the off-shell region below the $Z$ pole around a di-lepton invariant mass of 60 GeV.
The ambiguity is present because LEP had a limited ability to scan this region due to the kinematics of radiative return.
Hadron colliders, in contrast, have no such limitation and can probe the entire invariant mass range above and below the $Z$ pole.
In particular, as we will show, the $3~\abinv$ dataset of the high-luminosity run of the LHC (HL-LHC) is sufficient to conclusively break the sign degeneracy of $g_{b,R}$ at the $6\sigma$ level if the jet charge tagging efficiency is improved from 65\% to 80\% and by $4\sigma$ otherwise.

In this work, in order to sidestep the issue of overwhelming QCD backgrounds, we focus instead on $Z$ production in association with a $b$-quark pair with the $Z$ decaying to electrons and muons. Here, the background is much more manageable and, as we will show, dominantly arises from $t\bar t$ production where the top pair decays di-leptonically. A key feature of this background is that the di-leptonic final state is equally comprised of same- and different-flavor leptons. In contrast, the signal is of course only comprised of same-flavor leptons. This allows us to subtract the $t\bar t$ contribution at the expense of some statistical uncertainty which can be controlled by optimizing kinematic selection cuts to reduce this background as much as possible.

This paper is organized as follows.
In section~\ref{sec:theory}, we describe the theoretical framework, define the charge asymmetry observable, and describe its features including its dependence on the $Zbb$ couplings.
Then, in section~\ref{sec:simulation}, we discuss the signal and the composition of the background, describe the selection cuts, give details about the salient features of the analysis, and discuss the statistical, theoretical, and experimental systematic uncertainties and their treatment. In that section, we also identify di-leptonically decaying $\ttbar$ pair production as the largest background and discuss its subtraction using the different-flavor control region.
The results of this analysis are shown in section~\ref{sec:results}.
There, we compare the asymmetry without and with $\ttbar$ subtraction and show that the former is limited by the theory and the latter by statistics and thus benefits from larger datasets. To this end, we compare different integrated luminosity scenarios and show the effect of improving the charge-tagging efficiency.
Finally, we conclude in section~\ref{sec:conclusion} with a summary of our main findings, possible issues and future proposals.
Numerical fit results are documented in tables and figures in the appendices.

\section{Theoretical framework}
\label{sec:theory}

Deviations from the SM couplings of the fermions to the $Z$ boson can be parametrized as follows,
\begin{equation}
\lag_{Zff} = \sqrt{\frac{8G_F\,m_Z^2}{\sqrt{2}}}\,\sum_{\sigma} \left(g_{f,\sigma}^{\sm}+\delta g_{f,\sigma}\right)\bar{f}_\sigma \slashed{Z} f_\sigma\,,
\end{equation}
where the sum is over chiralities, $\sigma=L,R$ of the fermion, $f$. The coupling $g_{f,\sigma}^{\sm}$ is the SM one and includes loop corrections while $\delta g_{f,\sigma}$ only contains BSM contributions.
Thus, in the $\kappa$ framework, the modifier of the SM coupling is given by $\kappa_{f,\sigma}=1+\delta g_{f,\sigma}/g_{f,\sigma}^\sm$.
In the SM, the tree level left- and right-handed couplings are given by,
\begin{equation}
g_{f,L}^\sm = \left(T^3_f-\swsq\, Q_f\right)\,,\quad 
g_{f,R}^\sm = -\swsq\,Q_f\,,
\label{eq:gfr:sm}
\end{equation}
where $G_F$ is the Fermi constant, $\thw$ is the weak mixing angle, and $T^3_f=\nicefrac{\pm 1}{2}$ and $Q_f$ are the weak isospin and electric charge of the fermion, respectively.
In this work, our focus is on the coupling of the $Z$ to $b$-quarks and the deviations from the SM couplings $\delta g_{b,\sigma}$ can, for example, be generated by dimension-six operators in the Standard Model effective field theory (SMEFT)~\cite{Buchmuller:1985jz,Grzadkowski:2010es}.
In the Warsaw basis~\cite{Grzadkowski:2010es}, the relevant terms in the effective Lagrangian are,
\begin{equation}
\lag\supset 
 c_{\varphi {q,i}}^{(1)}\,(\overline{Q}_i  \gamma^{\mu} Q_i)(i H^{\dagger}  \overset{\leftrightarrow}{D}_{\mu} H)
 + c_{\varphi {q,i}}^{(3)}\,(\overline{Q}_i \sigma^{a} \gamma^{\mu} Q_i)(i H^{\dagger} \overset{\leftrightarrow}{D}{}^a_{\mu} H)
 + c_{\varphi {d,i}}\,(\overline{d_i}\gamma^{\mu} d_i)(i H^{\dagger} \overset{\leftrightarrow}{D}_{\mu} H)\,,
\end{equation}
where $Q_i$ is the left-handed $SU(2)$ quark doublet and $d_i$ is the right-handed down-quark. In both cases, the subscript $i$ denotes the quark flavor. Note that we only consider flavor-diagonal operators in the mass eigenbasis.
The left-right-acting derivatives are defined in the usual way as $\overset{\leftrightarrow}{D}_{\mu}\equiv \overset{\rightarrow}{D}_\mu-\big[\overset{\leftarrow}{D}_\mu\big]^\dagger$ and $\overset{\leftrightarrow}{D}{}^a_{\mu}\equiv 
\sigma^a\overset{\rightarrow}{D}_\mu-\big[\overset{\leftarrow}{D}_\mu\big]^\dagger\sigma^a$, where $\sigma^a$ are the Pauli matrices.
Note that the Wilson coefficients, $c_i$, are dimensionful via their dependence on the SMEFT scale, $\Lambda$ which can be interpreted as the scale at which new particles generate the effective operators.
With these conventions, the induced deviations in the $\zbb$ couplings read~\cite{deFlorian:2016spz}, 
\begin{equation}
\delta g_{b,L} = -\frac{1}{2\sqrt{2}G_F}\left(c_{\varphi {q,3}}^{(1)}+c_{\varphi {q,3}}^{(3)}\right)\,,\quad
\delta g_{b,R} = -\frac{1}{2\sqrt{2}G_F}\,c_{\varphi {d,3}}\,.
\end{equation}
In addition, there are flavor-universal contributions that arise from field redefinitions of the gauge bosons that are necessary to canonically normalize their kinetic terms.
These contributions are given in the Warsaw basis in appendix A of ref.~\cite{Breso-Pla:2021qoe} and we omit them here since they are better constrained by other observables, see for example~\cite{Breso-Pla:2021qoe}.
Furthermore, contributions from dipole operators are suppressed by the $b$-quark mass at leading order in the $1/\Lambda$ expansion and can therefore be neglected.
Finally, four-fermion operators could interfere with the SM amplitude but this interference is suppressed by $\mll^2/\Lambda^2$ which is $\mathcal{O}(1\%)$ for $\Lambda=1$ TeV.

\begin{figure}[t]\centering
	\includegraphics{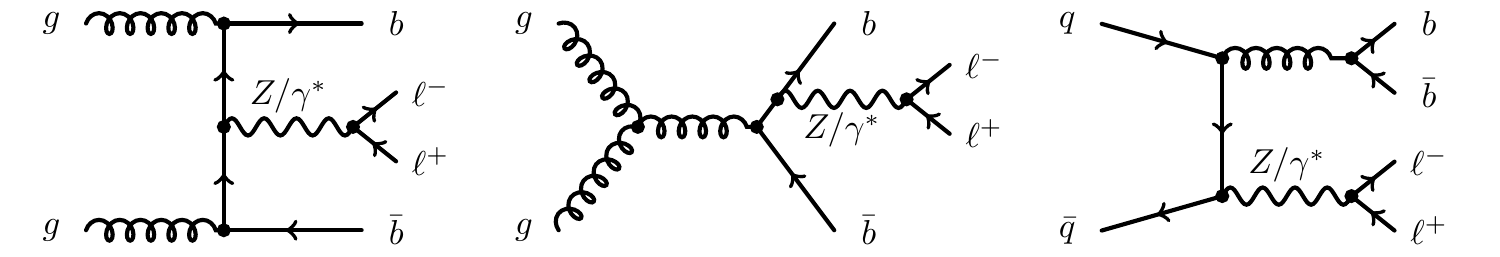}
	\caption{Representative Feynman diagrams for $pp\to b\bar{b}\ell^+\ell^-$. Here, $\ell$ represents an electron or a muon. In the rightmost diagram, the electroweak gauge bosons do not couple to the $b$-quarks and thus, this will contribute to the background, see text for more details.}
	\label{fig:feynman}
\end{figure}

The strongest constraints on the $\zbb$ couplings are derived from global fits to LEP measurements, i.e. precision electroweak measurements. They are mostly driven by the on-shell region where the cross-section depends on the squares of the couplings.
As mentioned in the introduction, we will leverage the fact that in hadron colliders there are sufficient events in the off-shell region. Specifically, the interference between the amplitudes mediated by the $Z$ boson and the photon affords us sensitivity to deviations that linearly depend on the $Z$ couplings. The process we focus on in this paper is $pp\to b\bar{b}\ell^+\ell^-$ where di-lepton pair in the final state is mediated by a $Z$ boson or a photon. Representative Feynman diagrams for this process are shown in figure~\ref{fig:feynman} where the left and middle diagrams represent signal processes while the right diagram represents a background process.
However, such a background that is due to gluon splitting can be reduced to a subdominant level as shown in the next section since the kinematics of the resulting $b$-quark pair are quite distinct from the signal.

\subsection{Coupling dependence of the squared amplitudes}
The coupling dependence of the signal diagrams (left and middle ones of figure~\ref{fig:feynman}) can be most easily understood if one considers the sub-amplitude $b\bar{b}\to Z/\gamma^*\to \ell^+\ell^-$.
In our analysis, we neglect the effect of the $b$-quark mass which contributes at the order of $\sim m_b^2/m_{\ell\ell}^2 \lesssim\mathcal{O}(1\%)$, see for example a recent calculation in ref. \cite{Bernreuther:2016ccf}. 
With this assumption, specifying the $b~(\ell^-)$ helicity fixes that of the $\bar{b}~(\ell^+)$ and the sub-amplitude of interest is given by,
\begin{equation}
\begin{split}
i\mcM_{\rho\sigma}=\parbox[c]{4cm}{\includegraphics{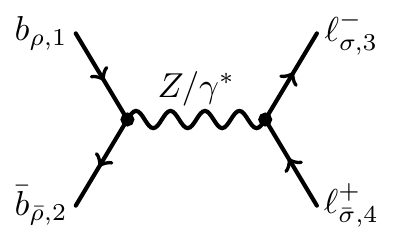}}&= i\,e^2 m_{\ell\ell}^2
\left(1+\frac{\rho}{\sigma}\cos\theta\right)
\left[
\frac{\,Q_\ell\,Q_b}{\propg} + 
\frac{g_{b,\rho}\,g_{\ell,\sigma}}{\cw^2\sw^2\propz}
\right]\,,
\end{split}
\label{eq:bbll_amplitude}
\end{equation}
where $\propg=\mll^2$ and $\propz=\mll^2-m_Z^2 + i m_Z\Gamma_Z$ are the photon and $Z$ propagator denominators, $\sw$ ($\cw$) is the sine (cosine) of the weak mixing angle.
The $Zff$ couplings, $g_{f,\rho}$, in the SM are given explicitly in eq.~\eqref{eq:gfr:sm}, $e$ is the gauge coupling of QED, and $Q_f$ is the electric charge of fermion $f$.
The helicities of the fermions, i.e., the $b$-quark and lepton $(\ell^-)$, are denoted by the labels $\rho,\sigma=\pm 1$.
Thus, the ratio $\rho/\sigma=\pm$, controls the sign in front of $\cos\theta$. The scattering angle, $\theta$, is defined as the angle between $\vec{p}_b-\vec{p}_{\bar{b}}$ and $\vec{p}_{\ell^-}-\vec{p}_{\ell^+}$ in the di-lepton rest frame, i.e.,
\begin{equation}
\cos\theta = \frac{(\vec{p}_b-\vec{p}_{\bar{b}})\cdot(\vec{p}_{\ell^-}-\vec{p}_{\ell^+})}{\lVert\vec{p}_b-\vec{p}_{\bar{b}}\rVert\,\lVert\vec{p}_{\ell^-}-\vec{p}_{\ell^+}\rVert}\,.
\label{eq:cos_theta_ll_rest_frame}
\end{equation}
Throughout the paper we use the $L/R$ and $\pm$ notations interchangeably where $L\equiv -$ and $R\equiv +$ as can be inferred from the sign that appears in the respective projection operator.
Crucially, eq.~\eqref{eq:bbll_amplitude} shows that the sign in front of the $\cos\theta$ term can only be negative when the $b$-quark and the lepton, $\ell^-$, have opposite helicities.
Furthermore, since under the exchange $\vec{p}_b\leftrightarrow\vec{p}_{\bar{b}}:~\theta\to\pi-\theta$ and so $\cos\theta\to-\cos\theta$.
That is, exchanging the quark and anti-quark momenta flips the sign of the $\cos\theta$ term.
This makes explicit the statement that charge conjugation which exchanges a particle with its anti-particle, equivalently flips the helicity of fermion line~\cite{Dixon:1996wi}.\footnote{Another way to see that the exchange $p_b\leftrightarrow p_{\bar{b}}$ is equivalent to flipping the helicities of both $b$ and $\bar{b}$ is as follows.  In the spinor helicity formalism, the amplitude in eq.~\eqref{eq:bbll_amplitude} is a product of two fermionic currents of the form $\langle i\gamma^\mu j]$. Exchanging the momenta, $i\leftrightarrow j$, exchanges the angle and square bracket and hence the helicities.}
Squaring the four helicity amplitudes, and summing over initial and final helicities gives,
\begin{equation}
\sum_{\rho,\sigma=\pm}\left|\mcM_{\rho\sigma}\right|^2=\left|\mcM_S\right|^2+\left|\mcM_A\right|^2\,,
\end{equation}
where the subscripts $S$ and $A$ stand for symmetric and antisymmetric, respectively, with respect to the exchange of the $b$ and $\bar{b}$ momenta. That is,
\begin{equation}
p_b\leftrightarrow p_{\bar{b}}\;:\;
\begin{cases}
\left|\mcM_S\right|^2 \to +\left|\mcM_S\right|^2 \\
\left|\mcM_A\right|^2 \to -\left|\mcM_A\right|^2
\end{cases} \,.
\label{eq:msquared_symmetry}
\end{equation}
And based on the arguments presented above, the symmetric and antisymmetric components of the squared matrix element in eq.~\eqref{eq:msquared_symmetry} take the form,
\begin{equation}
\begin{split}
\left|\mcM_S\right|^2 &= f_0 + f_1(g_L + g_R) + f_2 (g_L^2+g_R^2)\,,\\
\left|\mcM_A\right|^2 &=  h_1 (g_L - g_R) + h_2 (g_L^2-g_R^2)\,.\\
\end{split}
\label{eq:msquared_comps}
\end{equation}
Here, the $f_i$ and $h_i$ are functions of kinematic variables and lepton (but not $b$-quark) couplings.
For the full $pp\to \bbll$ squared amplitudes, the form of eq.~\eqref{eq:msquared_comps} still holds, of course, because this is a generic decomposition.
What is important, moreover, is that the property in eq.~\eqref{eq:msquared_symmetry} also holds because it follows from the action of charge conjugation along the $b$-quark line.

\subsection{Charge asymmetry}
\label{sec:charge_asymmetry}

\begin{figure}[t]\centering
	\includegraphics[scale=1.5]{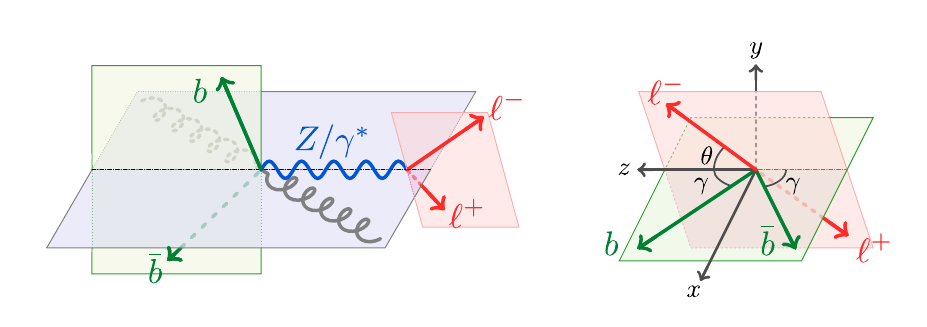}
	\caption{The left figure shows a kinematic configuration in the center-of-mass frame. The right figure shows the $bb\ell\ell$ system in the dilepton rest frame after an $O(3)$ rotation to align the $b$ and $\bar{b}$ momenta so that they both make an angle $\gamma$ with the $\pm \hat{z}$ direction; this is equivalent to the Collins-Soper frame for Drell-Yan.}
	\label{fig:cs_angle_ll_frame}
\end{figure}

A Lorentz invariant form of the dot product in eq.~\eqref{eq:cos_theta_ll_rest_frame} is, 
\begin{equation}
(p_b-p_{\bar{b}})\cdot(p_{\ell^-}-p_{\ell^+})\,,
\label{eq:asymmetric_observable}
\end{equation}
which, in the rest frame of the di-lepton system, leads to
\begin{equation}
-(\vec{p}_b-\vec{p}_{\bar{b}})\cdot(\vec{p}_{\ell^-}-\vec{p}_{\ell^+})
=-\lVert\vec{p}_b-\vec{p}_{\bar{b}}\rVert\,\lVert\vec{p}_{\ell^-}-\vec{p}_{\ell^+}\rVert\cos\theta\,.
\label{eq:cs_angle_ll_frame}
\end{equation} 
This definition is equivalent to the usual Collins-Soper~\cite{Collins:1977iv} angle for the lepton pair in the Drell-Yan process up to an $O(3)$ rotation which leaves the Euclidean dot product, and thus the definition of $\theta$, invariant in that frame, \- see figure~\ref{fig:cs_angle_ll_frame} and also ref.~\cite{Gauld:2017tww} for a clear discussion of the Collins-Soper frame.
From eq.~\eqref{eq:asymmetric_observable} and the right-hand side of eq.~\eqref{eq:cs_angle_ll_frame} one can see that while the value of $\cos\theta$ is frame dependent, its sign is not; i.e., the sign is, as expected, Lorentz invariant.
This was verified via explicit computation of the squared matrix element of the full  $gg\to bb\ell\ell$ process using \texttt{FeynCalc}~\cite{Mertig:1990an}.

The charge asymmetry, $A$, is then defined in the usual way by taking the difference over the sum of the cross-sections with positive and negative $\cos\theta$, i.e.,
\begin{equation}
A = \frac{\sigma(\cos\theta>0)-\sigma(\cos\theta<0)}{\sigma(\cos\theta>0)+\sigma(\cos\theta<0}\equiv \frac{\sigma_>-\sigma_<}{\sigma_>+\sigma_<}.
\label{eq:charge_asymmetry}
\end{equation}

\section{Monte Carlo simulation and event selection}
\label{sec:simulation}

Since we are interested in the charge asymmetry of the $b$-quark with respect to the direction of the lepton system, cf. eq.~\eqref{eq:asymmetric_observable}, we need to identify two $b$-tagged jets as well as their charges, and require them to have opposite signs.
The final state, therefore, consists of two $b$-tagged jets with opposite charges and two same-flavor, opposite-sign, muons or electrons.
The backgrounds that could contribute to such a final state can be split into two categories. In the first, we have $t{\bar{t}}$, single top $(Wtb)$, and di-boson $(ZZ/Z\gamma^*)$ production, all decaying to the same visible final state particles. In the second we have fake backgrounds mostly from $Z/\gamma^*$ in association with light or $c$-flavored jets. Throughout the analysis, we assume prospective (mis-)identification efficiencies $\varepsilon(q\to b)$, $q=j,c,b$ where $j$ denotes a light jet, i.e, $j=g,u,d,s$ while $c$ and $b$ denote a jet associated to a parton-level $c$ or $b$ quark. The working points are taken as $\varepsilon(j\to b) = 1\%$ and $\varepsilon(c\to b) = 10\%$ and a $b$-tagging efficiency $\varepsilon(b\to b) = 85\%$. 

\subsection{Signal and backgrounds}

\begin{figure}[t]
	\centering
	\begin{minipage}{.28\textwidth}
		\centering
		\includegraphics[width=0.98\linewidth]{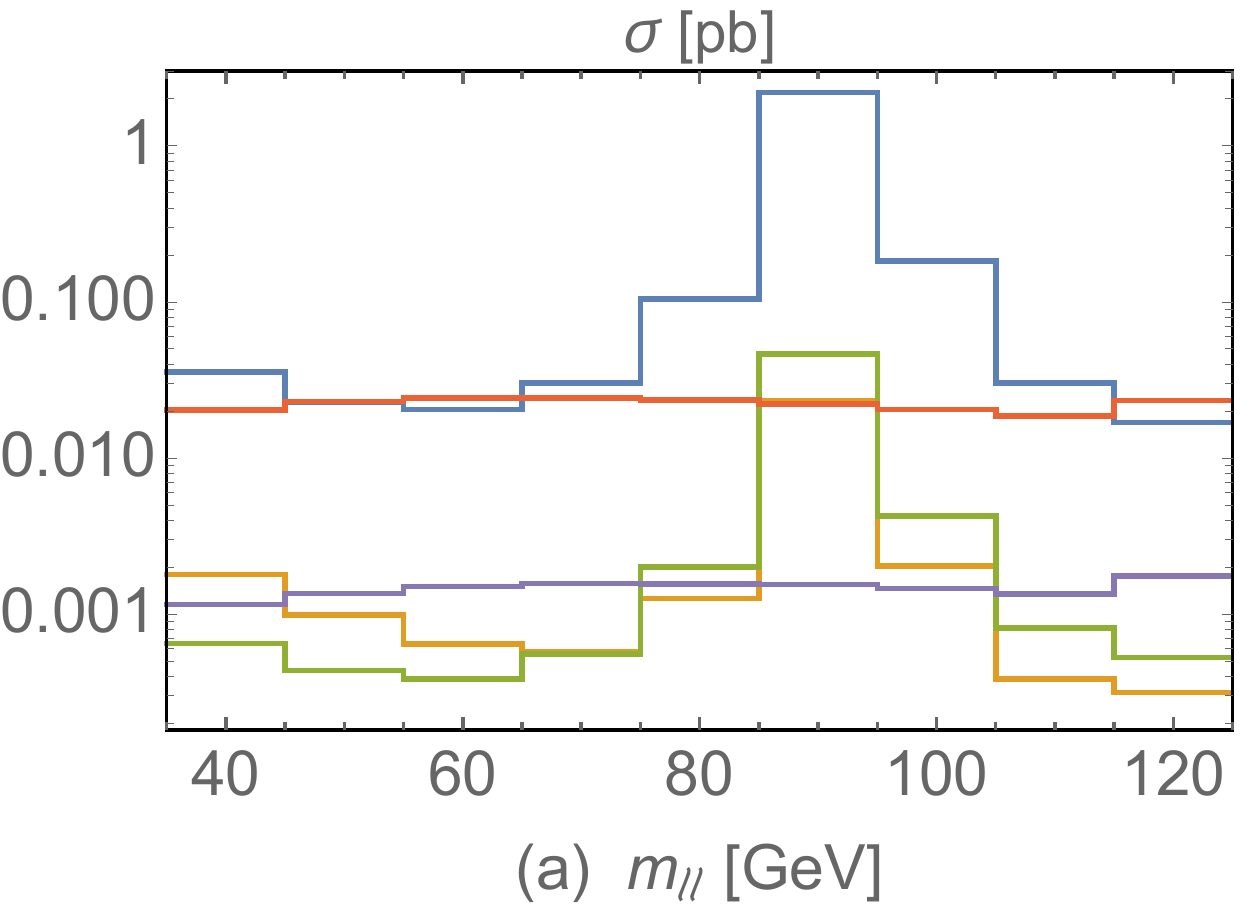}
	\end{minipage}%
	\begin{minipage}{.27\textwidth}
		\centering
		\includegraphics[width=0.98\linewidth]{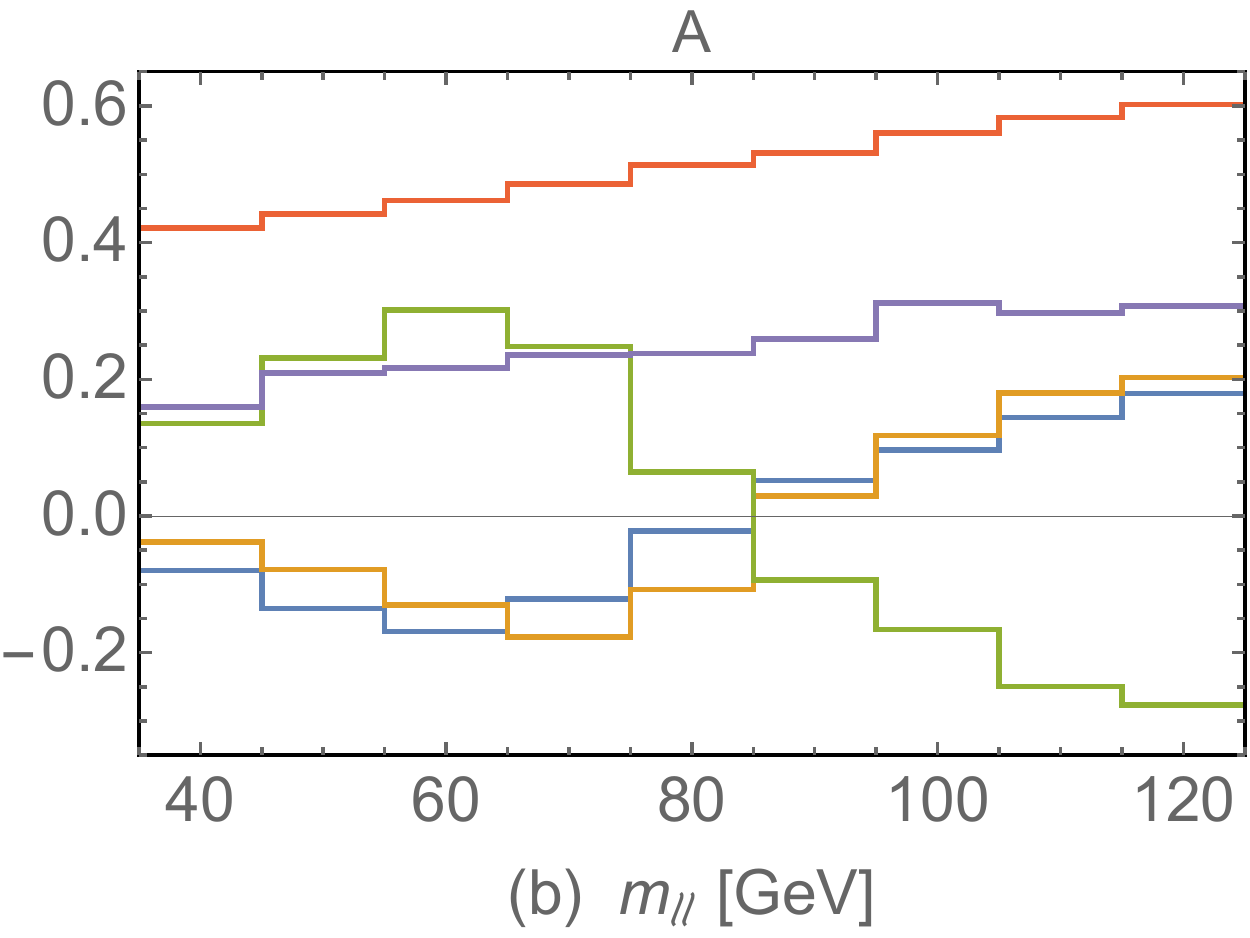}
	\end{minipage}%
	\begin{minipage}{.45\textwidth}
		\centering
		\includegraphics[width=0.98\linewidth]{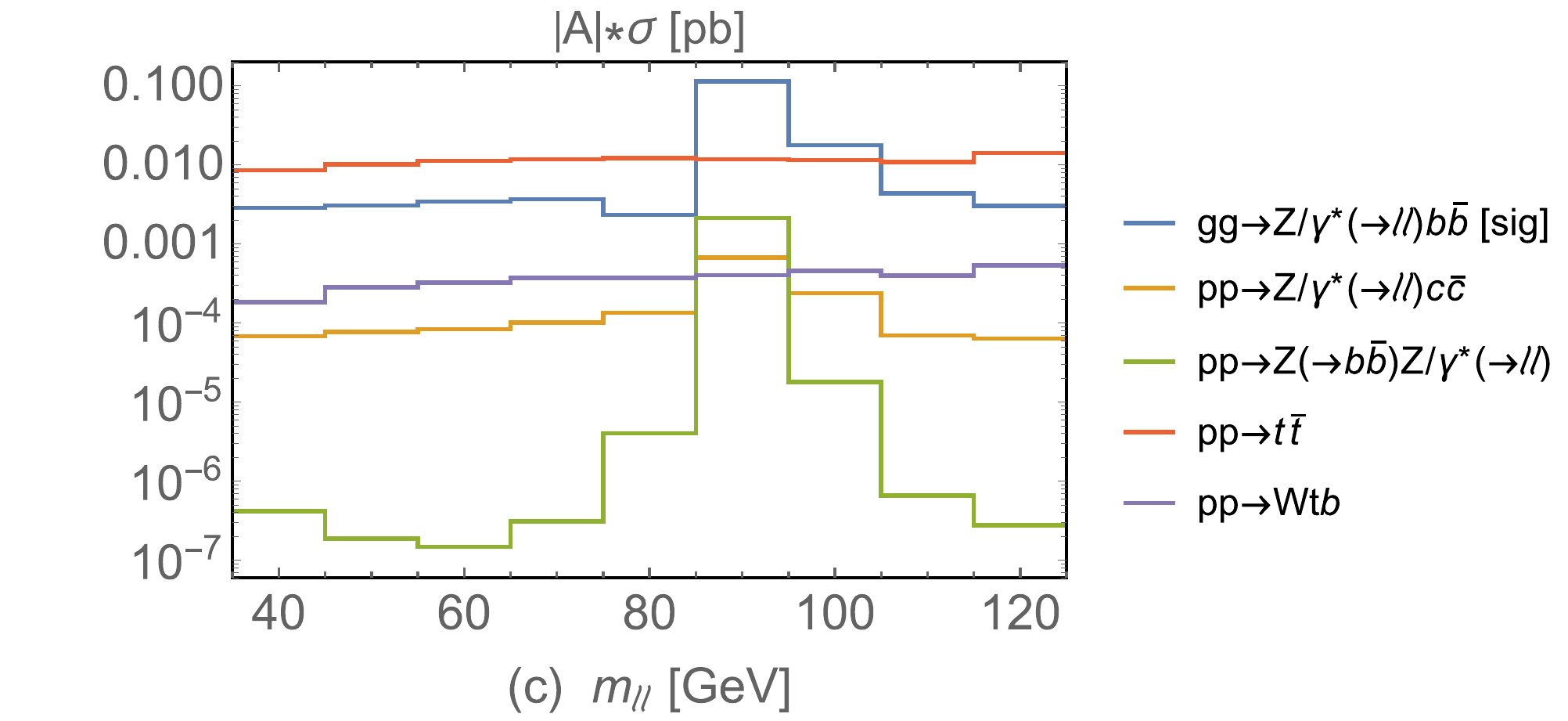}
	\end{minipage}%
	\caption{
	The cross section (a), asymmetry (b), and absolute value of the asymmetric cross-section (c) in the 10-GeV-wide bins we use in our analysis and after the selection cuts defined in eq.~\eqref{eq:final_acceptance_cuts} are applied.
	In all three panels, we show the signal and background contributions listed in eq.~\eqref{eq:signal_and_backgrounds}.
	}
	\label{fig:bkg_parton}
\end{figure}

To survey the contributions of the signal and the backgrounds discussed above, we first generate parton-level samples in \madgraph~\cite{Alwall:2014hca} at leading order (LO) using the following final selection cuts,\footnote{The lepton $p_T$ and $\eta$ cuts are set to be (almost) as inclusive as the trigger requirement allows. As a reference, we follow the current ATLAS $Vh$ analysis into the same final state~\cite{ATLAS:2020fcp}.}
\begin{equation}
\begin{split}
p_T^b > 20~\gev,\quad
p_T^\ell > 10~\gev,\quad
|\eta_{b,\ell}| < 2.5,\quad
\Delta R_{bb,\ell\ell, b\ell} > 0.4,\quad
\slashed{E}_T < 30~\gev\,.
\label{eq:final_acceptance_cuts}
\end{split}
\end{equation}
Furthermore, we consider di-lepton invariant masses in the range,
\begin{equation}
35 < \mll < 125~\gev\,,
\label{eq:mll_cut}
\end{equation}
in nine, 10 GeV bins.
The strong missing transverse energy ($\met)$ cut in eq.~\eqref{eq:final_acceptance_cuts} is necessary to reduce the $t\bar t$ background due to the presence of neutrinos.
For a center-of-mass energy $\sqrt{s}=14$ TeV and after applying the cuts of eqs.~\eqref{eq:final_acceptance_cuts} and~\eqref{eq:mll_cut} and multiplying by the $b$-tagging efficiencies, the signal and the backgrounds have the following cross-sections.
\begin{equation}
	\begin{tabular}{c c}\toprule[1pt]
Process & $\sigma\times\varepsilon(q\to b)^2$ [pb]\\\midrule
$Z/\gamma^*(\to\ell\ell)+bb$ & 2.2 \\
 $Z/\gamma^*(\to \ell\ell)Z(\to bb)$ & $5.5\times 10^{-2}$ \\
$tt\to bb\ell\ell\nu\nu$ (SF)& 0.33 \\
$Wtb\to bb\ell\ell\nu\nu$ (SF)& $1.1\times 10^{-2}$ \\
$Z/\gamma^*(\to\ell\ell)+cc$ & $3.0\times 10^{-2}$ \\
 $Z/\gamma^*(\to\ell\ell)+jj$ & $3.2\times 10^{-2}$\\
 $Z/\gamma^*(\to\ell\ell)+jc$ & $1.6\times 10^{-2}$\\\bottomrule[1pt]
\end{tabular}
\label{eq:signal_and_backgrounds}
\end{equation}
The differential cross-sections with respect to the di-lepton invariant mass, $d\sigma/dm_{\ell\ell}$, the charge asymmetry, $A$, as well as the asymmetric component of the cross-section, $\sigma_A$, for these contributions are shown in figure~\ref{fig:bkg_parton}.
For the various backgrounds considered here, the largest one by far is $t\bar t$ over the entire $\mll$ range. We will thus focus on the signal and $t\bar t$ processes in the full simulation and relegate the treatment of the other backgrounds to a more detailed study where other sub-leading effects could also be taken into account.

\subsection{Event generation}

Computing the charge asymmetry in eq.~\eqref{eq:charge_asymmetry} suffers from  the `loss of significance' problem, at least for the signal process, since it involves a difference of two quantities that are of the same numerical order, namely, $\sigg$ and $\sigl$.
Therefore, in each $\mll$ bin in our analysis, we need to compute these cross-sections with a very small Monte Carlo statistical uncertainty.
To this end, we generated 10 million events in each bin for the SM (signal and $t\bar t$ background) and for each additional $(\delta g_L,\delta g_R)$ benchmark as needed to construct the likelihood function.
This translates roughly to a total of $10^9$ events which was computationally time consuming.
Furthermore, the generation bins cover a wider of range of di-lepton invariant masses than the final analysis ones to mitigate against $\mll$ migration due parton shower effects; the generation range is, namely, $\mll\in[30,130]$ GeV.

To overcome the computational challenge, we generated events at leading order in QCD using the \texttt{gridpack} mode of \madgraph~\cite{Alwall:2014hca} which allowed us to make efficient use of available CPU cores.
Parton-level events with loose generation-level cuts were showered and hadronized with \pythia~\cite{Sjostrand:2014zea} and clustered with \fastjet~\cite{Cacciari:2011ma} (we used \hepmc~\cite{Bothmann:2022pwf} to pass events from \pythia to \fastjet on the fly).
We opted for the five-flavor scheme after checking explicitly that both the four- and five-flavor schemes lead to differential cross-sections and asymmetries that are compatible with each other at LO.
To account for next-to-leading-order (NLO) QCD corrections, we computed the total cross-section in each $\mll$ bin for the signal (SM point only) and the background in \mcfm~\cite{Campbell:2015qma} at LO and NLO.
We then scaled the LO cross-sections, $\sigg$ and $\sigl$, with the computed bin-by-bin $K$-factor which turns out not to depend on $\mll$ and is $1.6$ for the signal and $1.8$ for the $\ttbar$ background.
These $K$-factors are consistent with those found in ref.~\cite{Ahrens:2011px} for $t\bar t$ and ref.~\cite{FebresCordero:2009xzo} for $\bbbarz$.
Furthermore, we verified that events produced using this procedure have kinematic distributions that are in very good agreement with those obtained by additionally performing a detector simulation using \delphes~\cite{deFavereau:2013fsa}.

\paragraph{Electroweak input parameters.} We take the input parameters in the on-shell scheme~\cite{Denner:1991kt}
where the weak mixing angle is a derived quantity and is defined by~\cite{Sirlin:1980nh}
\begin{equation}
\swsq \equiv \sw^2 \equiv 1-\frac{M_W^2}{M_Z^2}\,.
\end{equation}
The input parameters are: $M_W=80.379$~GeV, $\Gamma_W=2.085$~GeV, $M_Z=91.1876$~GeV, $\Gamma_Z=2.4952$~GeV, and $G_F=1.1663787\times 10^{-5}$~GeV$^{-2}$~\cite{Zyla:2020zbs}.

\paragraph{Jet clustering parameters and $b$-tagging.}
The jets are clustered using the anti-$k_t$ algorithm~\cite{Cacciari:2008gp} with a radius parameter $R=0.4$. To identify $b$-jets after hadronization, we loop through the jet constituents with $p_T>5$ GeV and match a final $B$ hadron (i.e., just before it decays) to the $b$-quarks from the hard event. A jet receives a $b$-tag if it contains a $B$ hadron within $\Delta R<0.2$ of a hard $b$ quark which also allows us to assign the charge of the jet.

\paragraph{Lepton isolation.}
Leptons in the final state must be sufficiently isolated from hadronic and other electromagnetic activity.
To this end, we follow the isolation procedure in ref.~\cite{deFavereau:2013fsa} which defines an isolation parameter as follows,
\begin{equation}
I(\ell) = \frac{\sum_{i\neq \ell}\,p_{T,i}}{p_{T,\ell}},\qquad\left\{i:\Delta R_{i\ell}<R_{\max{}}~\textrm{and}~p_{T,i}>p_{T,\min{}}\right\}\,.
\end{equation}
Following the \delphes HL-LHC card, we require $p_{T,\min{}}=0.1$ GeV and $\Delta R_{\max{}} = 0.3$ with $I(\ell=e)<0.1$ and $I(\ell=\mu)<0.2$.

\paragraph{BDT analysis.}
\label{sec:bdt}
To further reduce the $t\bar t$ background, two methods are used and compared, a cut and count analysis with $\slashed{E}_T<30$ GeV and a boosted decision tree (BDT) classifier trained on 16 kinematic variables, none of which depend on the charge of the $b$ quark. In this way, the BDT treats both $b$ and $\bar{b}$ equally and, thus, leaves the asymmetry unaffected.
As in the cut and count analysis, the missing transverse energy variable is the most important discriminant between the signal and the $t\bar t$ background. The value of the BDT cut is chosen such that it maximizes the signal significance, $S/\sqrt{S+B}$. A cut flow comparison between the conventional $\met$ and BDT analyses in three wide bins (to allow for sufficient statistics for in the BDT analysis) is shown in table~\ref{tab:xsec_lo}.
As seen from the table, the statistical significance for the signal cross section is improved from the cut-and-count analysis by about 20--30\% in the two off-shell regions where the BDT classifier can make use of the correlations and non-trivial boundaries in the 16-dimensional BDT input space. In the $Z$-pole region, however, the improvement is approximately 5\% since the $\ttbar$ background contributes sub-dominantly to the signal.
For the full analysis in section~\ref{sec:results}, we trained a BDT classifier in each of the nine bins that enter the likelihood function.
\begin{table}[H]
    \centering
    \begin{tabular}{c|c c c c|c c c c }
    \toprule[1pt]
         $\sigma$ (fb) & $b\bar b\ell\ell$ & 35--85 & 85--95 & 95--125 & $t\bar t$ & 35--85 & 85--95 & 95--125 \\\midrule
         Selection w/o $\met$
         & 700 & 72.9 & 565 & 62.1 & 1312 & 802 & 151 & 359 \\ 
         $\slashed{E}_T<30$ GeV & 605 & 63.9 & 488 & 53.2 & 225 & 132 & 26.3 & 67.1\\ 
         BDT cut & $632$ & 51.9 & 548 & 46.5 & $156$ & 24.8 & 43.2 & 18.4 \\ \bottomrule[1pt]
    \end{tabular}
    \caption{A simple cut-flow analysis of the leading-order cross-section for the signal and $\ttbar$ background. The BDT cuts are optimized to maximize  $S/\sqrt{S+B}$ assuming a luminosity of $3~\abinv$. Since a different BDT is trained and optimized in each of the three bins in the table, the cross-sections in these bins after the cut is applied do not sum to the total number in the corresponding leading column.}
    \label{tab:xsec_lo}
\end{table}

\subsection{The total charge asymmetry}
\label{sec:total_asymmetry}

When both top quarks decay to same-flavor (SF) charged leptons, the visible particles in the final state are identical to those of the signal process. As discussed above, this contribution can be reduced by cutting on the missing transverse energy. Moreover, it can be removed as will be discussed in the next subsection. Nevertheless, we discuss it briefly here because it does contribute to the charge asymmetry of the $b$-quarks due to the $V-A$ nature of the weak decay of the top. Note also that the charge asymmetry does not receive any contribution from QCD corrections and is unrelated to the $\ttbar$ forward-backward asymmetry for which the QCD contribution is the dominant one, for a review see~\cite{Kuhn:2011ri}.

The total number of observed events is,
\begin{equation}
    N = N_{\bbbarz}+N_{\ttbar}=\mathcal{L}\left(\sbbz+\stt\right)\,,
\end{equation}
where $N_i$ is the expected number of events in channel $i$ for an integrated luminosity $\mathcal{L}$ and $\sigma_i$ are the corresponding cross-sections. The total charge asymmetry (see eq.~\eqref{eq:charge_asymmetry}) receives contributions from both the signal and the $\ttbar$ background; it can be written as, 
\begin{equation}
    A= \frac{A_{\ttbar}\,\stt + A_{\bbbarz}\,\stt}{\sbbz + \stt}\,,
    \label{eq:total_asymmetry}
\end{equation}
which, for convenience, is given in terms of the individual asymmetries, $A_i$, and the cross-sections $\sigma_i$. The statistical uncertainty on this asymmetry is given by,
\begin{equation}
A_\text{stat} = \sqrt{\frac{1-A^2}{N_{\bbbarz}+N_{\ttbar}}}\,.
\label{eq:uncertainty_on_total_asymmetry}
\end{equation}

The total asymmetry is also susceptible to BSM contributions to the $\ttbar$ process itself.
Generically, these cancel in the subtracted asymmetry even in the presence of lepton-flavor non-universality, at least to leading order in the $1/\Lambda$ expansion where $\Lambda$ is the SMEFT scale. That is, if one neglects double-insertions of effective operators.

\subsection{Subtracting the $t\bar{t}$ background}
\label{sec:subtracting_ttbar}

Since the signal consists of SF leptons only, we can use the DF control region~\cite{ATLAS:2018kot,ATLAS:2020fcp} to subtract the $t\bar{t}$ contribution.
The subtracted cross-section is given by,
\begin{equation}
\bar \sigma = \sigma - \stt^\textsc{df}
= \sbbz + \stt^\saf -\stt^\dif \approx \sbbz \,.
\end{equation}
Of course, it still contributes to the statistical error which is now given by,
\begin{equation}
\mathcal{L}^2\,(\delta\bar{\sigma})^2=\delta\bar N^2 = \delta N^2 +  \left(\delta N_{\ttbar}^\dif\right)^2 = N_{\bbbarz} +2N_{\ttbar}\,.
\end{equation}
We assume that the theory uncertainties (due to missing higher orders) for the SF and DF contributions from $\ttbar$ are fully correlated and therefore mostly cancel.
Correspondingly, the subtracted charge asymmetry in eq.~\eqref{eq:charge_asymmetry} becomes,
\begin{equation}
\bar A = \frac{(\sigg-\sigg^\dif)-(\sigl-\sigl^\dif)}{(\sigg-\sigg^\dif)+(\sigl-\sigl^\dif)} \approx A_{bbZ}\,.
\end{equation}
Since the statistical uncertainties on the cross-sections, $\sigg$ and $\sigl$, 
are uncorrelated, the induced uncertainty on the subtracted charge asymmetry is then,
\begin{equation}
\delta\bar{A}_\text{stat}
=\frac{\sqrt{\sbbz\,\left(1-A_{\bbbarz}^2\right)+2\stt^\dif\,\left(1+A_\bbbarz^2-2 A_{\ttbar}^\dif A_{\bbbarz}\right)}}{\sqrt{\mathcal{L}}\,\sbbz}\,.
\label{eq:statistical_uncertainty_on_subtracted_asymmetry}
\end{equation}

\subsection{Theory and experimental uncertainties}
\label{sec:sys_error}

Apart from the statistical uncertainty on the total and subtracted asymmetries in eqs.~\eqref{eq:uncertainty_on_total_asymmetry} and~\eqref{eq:statistical_uncertainty_on_subtracted_asymmetry}, we consider two additional sources of uncertainty. The first arises from the theoretical prediction of the cross-sections and asymmetries because the perturbative expansion is truncated at a fixed order. Hence, this is an estimate of the effect of missing higher orders in the expansion. The second arises from experimental and other systematic effects. These two sources of uncertainty have different characters and we thus treat them separately.
 
\subsubsection{Theory uncertainty}
\label{sec:theory_uncertainty}

To estimate this uncertainty, we follow the standard procedure of varying the renormalization scale, $\mu_R$, and factorization scale, $\mu_F$ from a common central value, $\mu_0$, by a factor of $1/2$ and $2$ with the constraint $\tfrac{1}{2}\leq \mu_R/\mu_F\leq 2$.
We then take half of the envelope of the remaining seven scale combinations as a measure of the theory uncertainty.
In the case of $\ttbar$, the central value for both scales is set to~\cite{Czakon:2019txp,Mazzitelli:2021mmm},
\begin{equation}
\mu_{0,\ttbar}=\frac{1}{4}H_T=\frac{1}{4}\left(\sqrt{p_{T,t}^2+m_t^2}+\sqrt{p_{T,\bar{t}}^2+m_{\bar{t}}^2}\right)\,,
\end{equation}
and for $\bbbarz$, it is~\cite{Gauld:2023zlv},
\begin{eqnarray}
\mu_{0,\bbbarz}=E_{T,\ell\ell}=\sqrt{\mll^2+p_{T,\ell\ell}^2}\,.
\end{eqnarray}
Note also that unlike the Drell-Yan cross-section which is $\mathcal{O}(\alpha_s^0)$ at leading order, the $\bbbarz$ and $\ttbar$ cross-sections are $\mathcal{O}(\alpha_s^2)$.
Therefore, the scale variation envelope should be a reliable measure of the theory uncertainty especially at NLO.

To compute the scale variation envelope on the asymmetry, we assume that the scales in the numerator are correlated with those in the denominator.
For the total asymmetry, both the $\bbbarz$ and $\ttbar$ asymmetries appear (see eq.~\eqref{eq:total_asymmetry}) and for each, the theory error is computed as we just described. However, we consider the theory error on the two asymmetries independently; that is, we consider them to be uncorrelated with one another.
Under these assumptions, the theory uncertainty on the total asymmetry can be approximated by,
\begin{equation}
\begin{aligned}
(\delta A)^2 \approx (A_{\bbbarz}-A_{\ttbar})^2\frac{\stt^2\,\sbbz^2}{(\stt+\sbbz)^4}
\left[
\frac{(\delta\sbbz)^2}{\sbbz^2}+
\frac{(\delta\stt)^2}{\stt^2}
\right] 
+\frac{(\sbbz\,\delta A_{\bbbarz})^2+(\stt\,\delta A_{\ttbar})^2}{(\sbbz+\stt)^2}\,, 
\end{aligned}
\label{eq:theory_uncertainty_on_total_asymmetry}
\end{equation}
where the $\ttbar$ cross-section is either the SF or DF one but not their sum.

\subsubsection{Experimental and other systematic uncertainties}
\label{sec:systematic_uncertainties}

By contrast, the assumption we made above that $\stt$ and $\sbbz$ (and implicitly $\sigg$ and $\sigl$) are uncorrelated does not hold for experimental systematic uncertainties which are expected to be highly correlated for these quantities.
For this reason, we neglect these uncertainties in our analysis.\footnote{For the total asymmetry however, a more careful treatment of the experimental systematic uncertainties is required since the degree of correlation between $b\bar{b}Z$ and $\ttbar$ quantities is likely not so high.}
Nevertheless, we discuss them here for completeness.

The systematic uncertainties on the inclusive cross-section for $Z+\geq 2~b$-jets in a recent ATLAS analysis with $36~\fbinv$~\cite{ATLAS:2020juj} and a CMS one with $137~\fbinv$~\cite{CMS:2021pcj} are dominated by the jet energy scale and resolution (JES/JER), b-tagging and mis-tagging, and luminosity contributions; they are listed in table~\ref{tab:systab}.
For the $\ttbar$ cross-section, both ATLAS~\cite{ATLAS:2023gsl} and CMS~\cite{CMS:2021vhb} have performed an analysis with the full Run 2 dataset (140 and 137$~\fbinv$ respectively). CMS also has a very recent analysis from Run 3 with $\sqrt{s}=13.6$ TeV~\cite{CMS:2023qyl}. The uncertainties quoted in all three $\ttbar$ analyses are also listed in table~\ref{tab:systab}.
\begin{table}[t]
	\centering
	\begin{tabular}{c | c c | c c c}\toprule[1pt]
		&\multicolumn{5}{c}{Uncertainties [\%]}\\\midrule
		&  \multicolumn{2}{c|}{$Zb\bar b$} & \multicolumn{3}{c}{$t\bar t$} \\
		Source & ATLAS~\cite{ATLAS:2020juj} & CMS~\cite{CMS:2021pcj} & ATLAS~\cite{ATLAS:2023gsl} & CMS~\cite{CMS:2021vhb} & CMS~\cite{CMS:2023qyl}\\\midrule
		Jet energy scale & 14 & 5.8 & 0.1 & 0.3--1.8 & 0.7\\ 
		$b$-tagging & 5.0 & 5.8 & 0.07 & $\sim 3$~\cite{CMS:2017wtu} & 1.1\\ 
		Luminosity & 2.9 & 1.6\% & 0.93 & 2.3--2.5 & 2.3\\ \bottomrule[1pt]
	\end{tabular}
	\caption{Experimental systematic uncertainties on the fiducial cross-sections for $Z~+\geq 2b-$jets and $t\bar t$ from most recent ATLAS and CMS analyses; see text for details. The last column is from the most recent CMS measurement at $\sqrt{s}=13.6$ TeV with $1.21~\fbinv$~\cite{CMS:2023qyl} from}
	\label{tab:systab}
\end{table}

\section{Results}
\label{sec:results}

The symmetric and asymmetric $\mll$ distributions were binned in the range $\mll\in[35,125]$ GeV with a bin width of 10 GeV. In each bin, we generated sufficient $(g_L,g_R)$ benchmarks to obtain a reliable fit to the functional form in eq.~\eqref{eq:msquared_comps},
\begin{equation}
\begin{split}
\frac{d\sigma}{d\mll} &= F_0 + F_1(g_L + g_R) + F_2 (g_L^2+g_R^2)\,,\\
\frac{d\sigma_A}{d\mll} &= H_1 (g_L - g_R) + H_2 (g_L^2-g_R^2)\,.
\end{split}
\label{eq:dsigma_dmll_coeffs}
\end{equation}
In the case where the $Z$ boson or the photon couple to a $b$-quark and a lepton simultaneously, these coefficients can be written in terms of two common factors, $F$ and $H$, that only depend on $\mll$ as shown in appendix~\ref{app:dsigma}. In general, however, diagrams like the right-most one in figure~\ref{fig:feynman} break this relation (though not for $H$). The fitted values of the $F_i$ and $H_i$ coefficients are given in appendix~\ref{app:glgrfit}.

Assuming Gaussian statistics and no correlation between bins, the log-likelihood function is given by,
\begin{equation}
-2\Delta\log{L}(g_L,g_R)=\Delta\chi^2(g_L,g_R)=\sum_{\textrm{bins}}
\frac{\left[A(g_L,g_R)-A(g_L^\textsc{sm},g_R^\textsc{sm})\right]^2}{\delta A^2}\,,
\end{equation}
where the error, $\delta A^2=\delta A_\textrm{stat.}^2+\delta A_\textrm{theo.}^2$, is the sum, in quadrature, of the statistical and the theory errors. The latter mainly accounts for the uncertainty in the prediction of the charge asymmetry due to missing higher order corrections in the perturbative expansion as discussed in section~\ref{sec:theory_uncertainty}.
We assume that the experimental systematic uncertainties on the asymmetry are sub-leading largely because their effects on the numerator and denominator of the asymmetry are expected to be highly correlated, see section~\ref{sec:systematic_uncertainties}.
Therefore, their detailed consideration is beyond the scope of this work.

Since the charge asymmetry depends on the correct identification of the charge of the $b$-tagged jets, we must account for the effect of mis-tagging. Denoting the efficiency of correctly tagging the charge of the $b$-jets by $\epchrg$ and requiring two oppositely charged $b$-tagged jets, the asymmetry is rescaled by a factor,
\begin{equation}
A\to \frac{2\epchrg-1}{1-2\epchrg+2\epchrg^2}A .
\label{eqn:charge-efficiency-factor}
\end{equation}
In deriving this factor, we assumed that the efficiency of mis-tagging and correctly tagging the jet charge sum to one, i.e. $\varepsilon(Q\to -Q)=1-\varepsilon(Q\to Q)$. This corresponds to the symmetric cut on the discriminant of the jet vertex tagger algorithm developed by ATLAS~\cite{ATLAS:2015jhb,ATLAS-CONF-2018-022}.
Note also that the charge tagging efficiency mildly depends on the $b$-tagging efficiency as can be seen in table 5 of ref.~\cite{ATLAS:2015jhb}.
In our analysis, we take $\epchrg=65\%$ as our benchmark; this reflects the current performance of the ATLAS jet vertex charge tagger. In addition, we take an optimistic value of 80\% that could hopefully be reached by the time the HL-LHC achieves its full integrated-luminosity target.
For $\epchrg=65\%$ (80\%), the asymmetry is rescaled by a factor of 0.55 (0.88). 

\begin{figure}[t]
	\centering
	\includegraphics[width=0.49\linewidth]{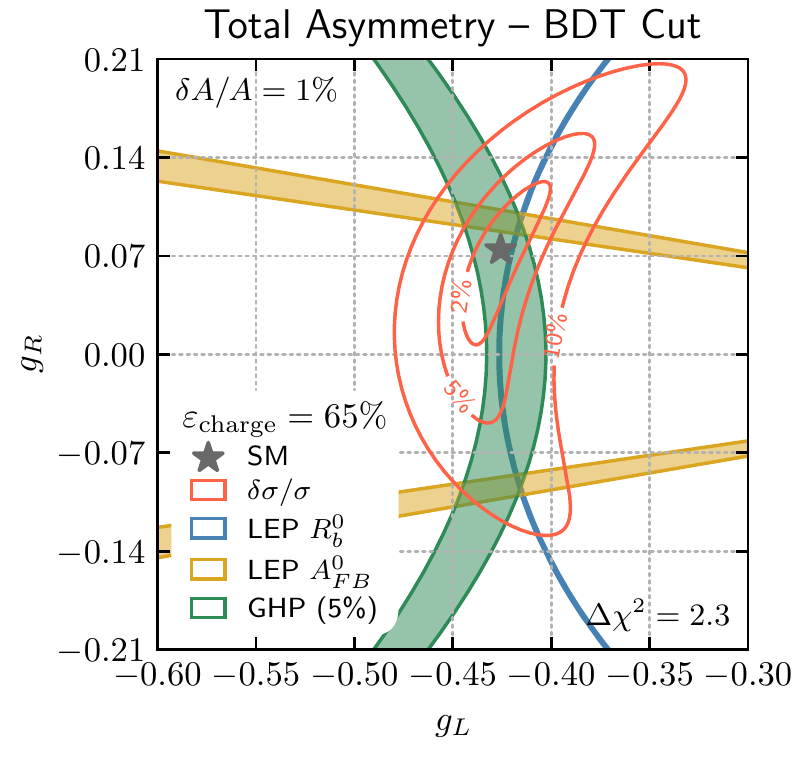}
	\includegraphics[width=0.49\linewidth]{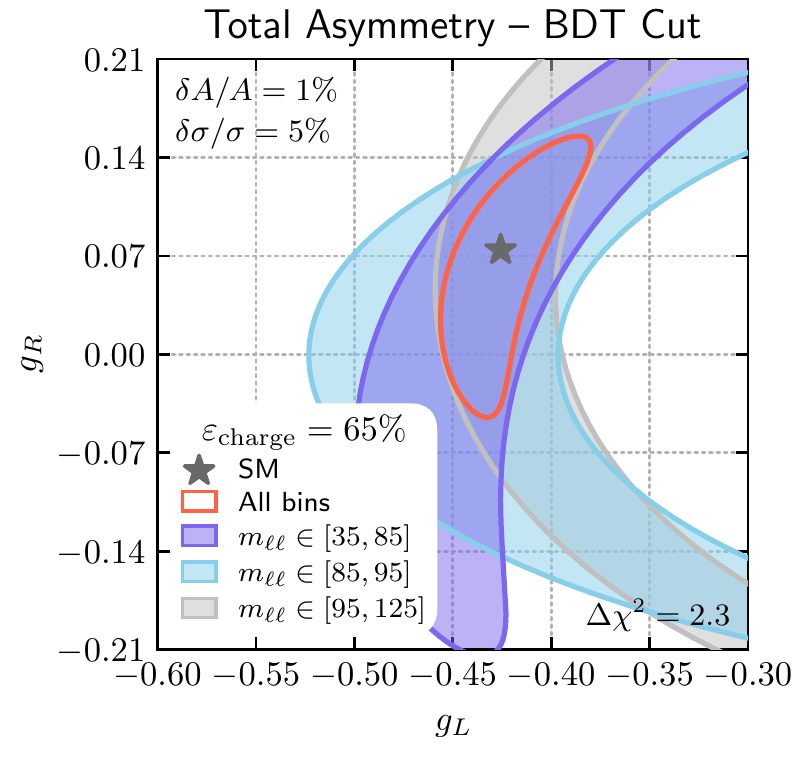}
	\caption{The $\Delta\chi^2=2.3$ (68\% CL) contours for the total asymmetry with the BDT cut. \textbf{Left panel:} combination of all nine $\mll$ bins; the bounds from LEP and projection at LHCb from ref.~\cite{Gauld:2019doc} with 5\% uncertainty (GHP) are shown for comparison.
	Only the theory error defined in eq.~\eqref{eq:theory_uncertainty_on_total_asymmetry} is included with $\delta\sigma/\sigma=\{2,5,10\}\%$ with fixed $\delta A/A=1\%$ for both $\ttbar$ and $bbZ$ in each bin. \textbf{Right panel:} individual contribution of the $\mll$ regions below, around, and above the $Z$ pole with $\delta\sigma/\sigma=5\%$.}
	\label{fig:total}
\end{figure}

\begin{figure}[t]
	\centering
	\includegraphics[width=0.49\linewidth]{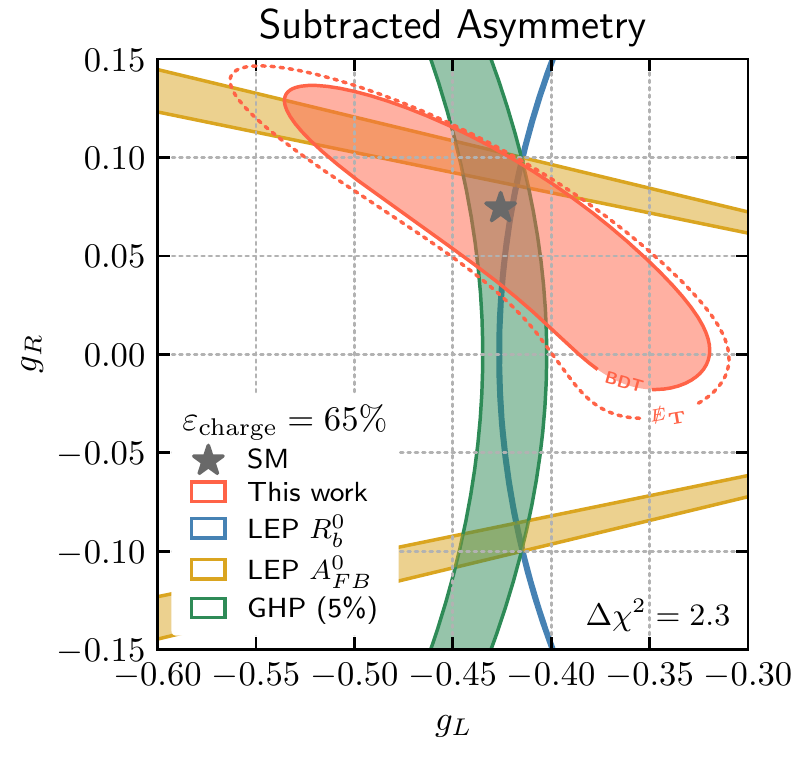}
	\includegraphics[width=0.49\linewidth]{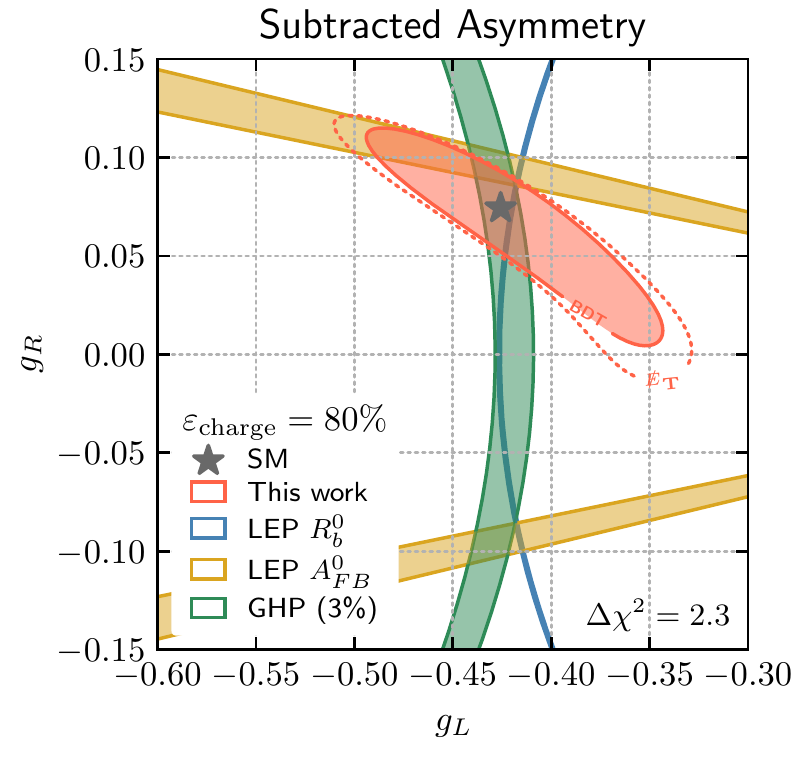}
	\caption{The $\Delta\chi^2=2.3$ (68\% CL) contours for the subtracted asymmetry. The solid (dashed) contour is obtained with the BDT ($\met$) cut. The LEP and LHCb (GHP) bands are described in the main text and the caption of figure~\ref{fig:total}.
	Two values of the charge tagging efficiency, $\epchrg$, are shown for comparison; a currently achievable one of 65\% in the left and a projected improvement of 80\% in the right panel.
}
	\label{fig:subtracted_combined}
\end{figure}

\begin{figure}[t]
	\centering
	\includegraphics[width=0.49\linewidth]{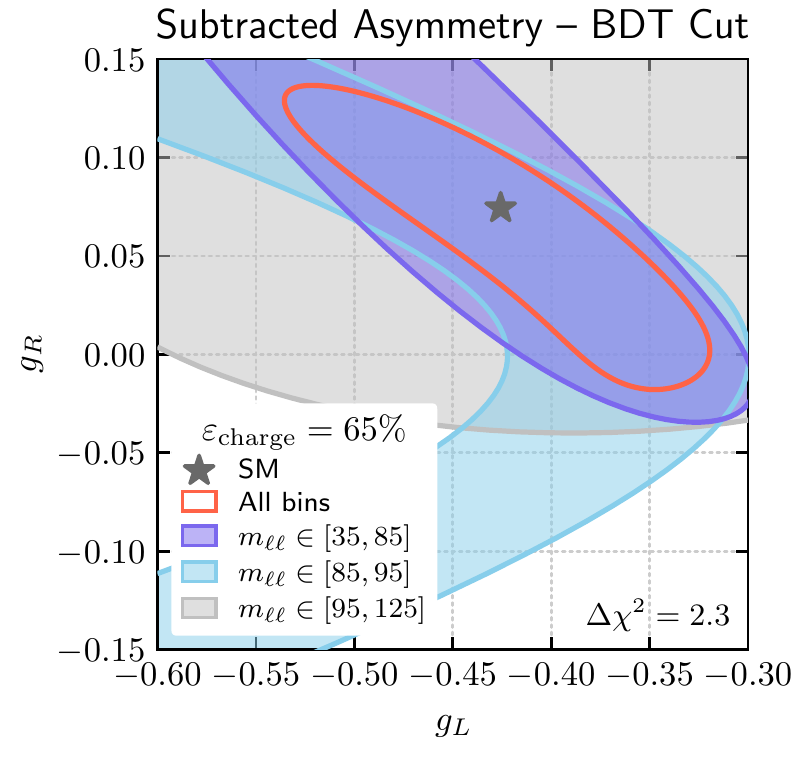}
	\includegraphics[width=0.49\linewidth]{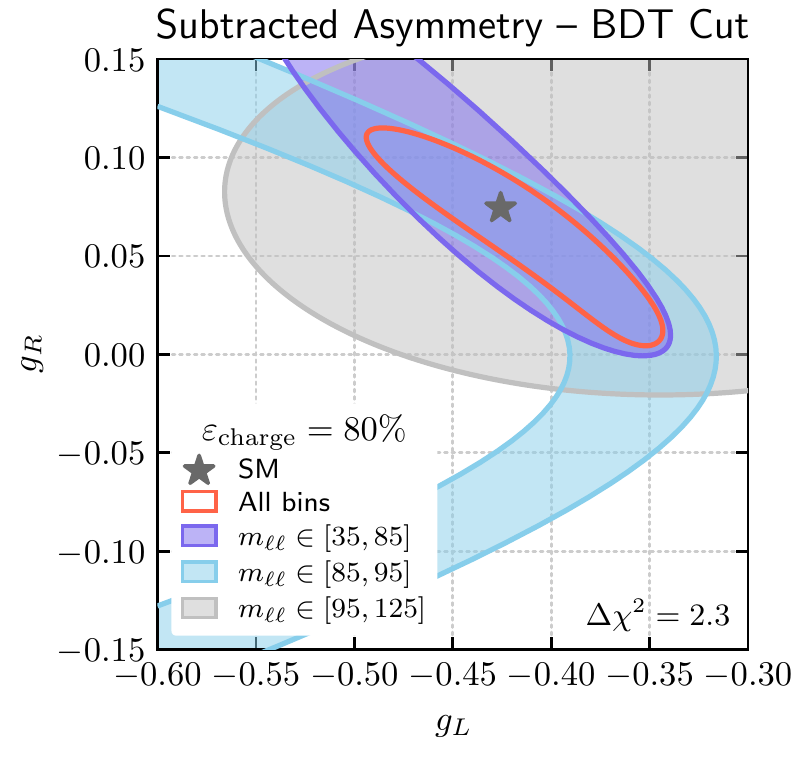}
	\caption{The $\Delta\chi^2=2.3$ (68\% CL) contours for the subtracted asymmetry with the BDT cut. 
	Three contours are shown for the $\mll$ regions below, around, and above the $Z$ pole; the exact ranges are listed in the legend. 
	Two values of the charge tagging efficiency, $\epchrg$, are shown for comparison; a currently achievable one of 65\% in the left and a projected improvement of 80\% in the right panel.}
	\label{fig:subtracted_bins}
\end{figure}

\begin{figure}[t]
	\centering
	\includegraphics[width=0.49\linewidth]{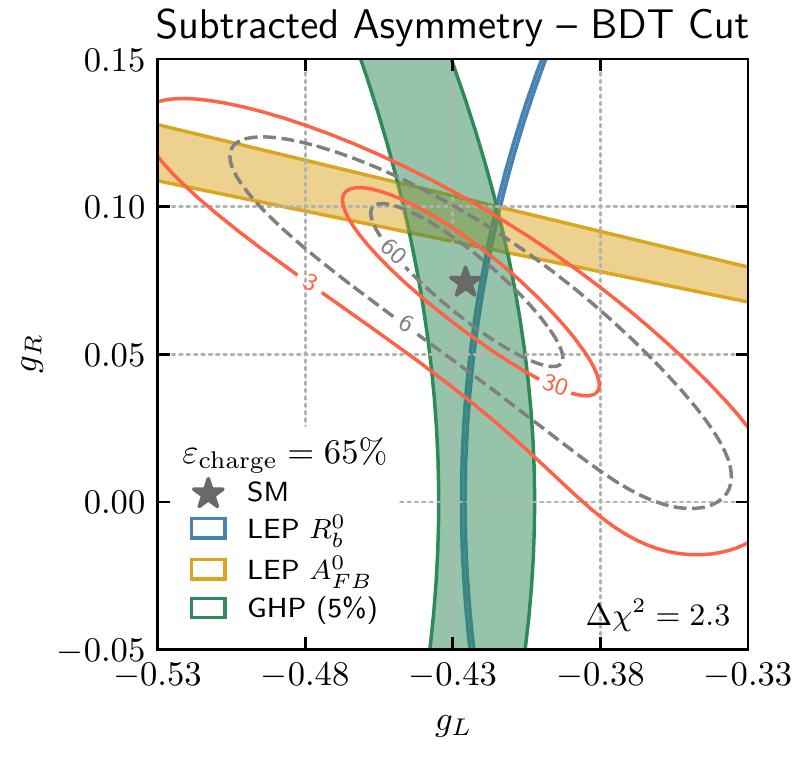}
	\includegraphics[width=0.49\linewidth]{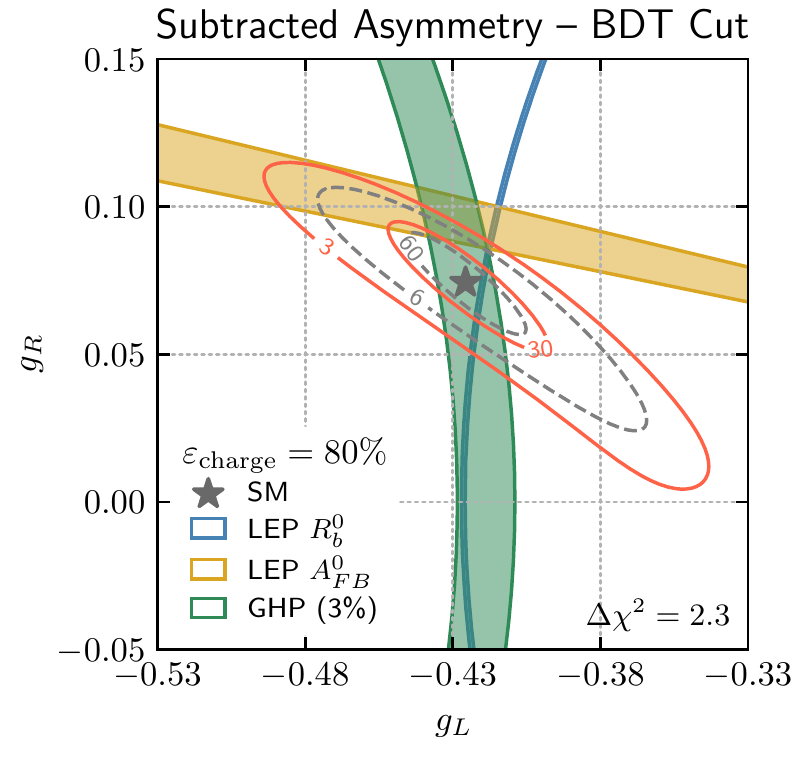}
	\caption{The $\Delta\chi^2=2.3$ (68\% CL) contours for the subtracted asymmetry with the BDT cut for various luminosity scenarios. 
	The solid red contours a baseline luminosity of 3 $\abinv$ and a projected one of 30 $\abinv$ corresponding to the full dataset of the HL-LHC and a possible FCC-hh, respectively.
	The dashed grey contours are a naive two-experiment combination obtained simply by doubling the two luminosities discussed above.
	Two values of the charge tagging efficiency, $\epchrg$, are shown for comparison; a currently achievable one of 65\% in the left and a projected improvement of 80\% in the right panel. The LEP and LHCb (GHP) bands are described in the main text and the caption of figure~\ref{fig:total}.}
	\label{fig:subtracted_lumi_projections}
\end{figure}

The HL-LHC is expected to achieve an integrated luminosity of $3~\abinv$~\cite{ZurbanoFernandez:2020cco}; this is the baseline, single-experiment, luminosity used to report our results in figures~\ref{fig:total},~\ref{fig:subtracted_combined}, and~\ref{fig:subtracted_bins}.
In addition, we explore three luminosity scenarios in figure~\ref{fig:subtracted_lumi_projections}. Namely, $6~\abinv$ for an ATLAS and CMS combination at the end of HL-LHC and 30 (60)~$\abinv$ for an estimate of the single-experiment (ATLAS and CMS combination) of the integrated luminosity target of the FCC. While we haven't preformed an FCC-specific study, our projection is likely a conservative estimate of the reach since, at the very least, the gluon luminosity will be higher at FCC with respect to the LHC for the same partonic center-of-mass energy due to the low-$x$  enhancement of the gluon parton distribution function~\cite{Arkani-Hamed:2015vfh}.

We also compare our bounds in figures~\ref{fig:total}, \ref{fig:subtracted_combined}, and~\ref{fig:subtracted_lumi_projections} to those on $R_b^0$ and $A_{FB}^{0,b}$ from LEP~\cite{ALEPH:2010aa} which are shown as blue and goldenrod bands, respectively. However, we use the updated value for $A_{FB}$ from~\cite{Bernreuther:2016ccf} where the NNLO QCD corrections with massive bottom quarks were computed. The values we use are,
\begin{equation}
\begin{split}
R_b^0 &= 0.21629\pm 0.00066\,,\\
A_{FB}^{0,b} &= 0.0996\pm 0.0016\,.
\end{split}
\end{equation}	
We also show the bounds from the proposal of Gauld, Haisch, and Pecjak (GHP)~\cite{Gauld:2019doc} at LHCb.
Specifically, we show the 3\% or 5\% uncertainty band, depending on the figure, in green for the ratio of the bottom to charm asymmetry $R_{b/c}$. 
This asymmetry is defined as the difference of the heavy quark and anti-quark rapidities in the (forward) LHCb fiducial region.
The dependence of the related observables on the deviations from the SM values, $\delta g_L$ and $\delta g_R$, is shown in appendix~\ref{sec:comparison_bounds} where the expressions are expanded to leading order to illustrate their linearized behavior near the SM point. Note that the authors of~\cite{Gauld:2019doc} extract the SM couplings from a fit; for this reason, we shift their bounds so that they are centred around the SM couplings computed in our electroweak input scheme.

The left panel of figure~\ref{fig:total} shows the $\Delta\chi^2=2.3\approx 68\%$ C.L. contours for a relative theoretical uncertainty of 2\%, 5\%, and 10\% on the cross-sections and with a fixed relative uncertainty on the asymmetries, $\delta A/A=1\%$, see eq.~\eqref{eq:theory_uncertainty_on_total_asymmetry}. The total, i.e. un-subtracted, asymmetry includes the contribution of the $t\bar{t}$ background process to the charge asymmetry.
And the right panel breaks down the contributions of the nine $\mll$ bins into three regions below, around, and above the $Z$ pole with an $\mll$ coverage of $[35,85]$, $[85,95]$, and $[95,125]$ GeV respectively. This breakdown illustrates the (unsurprising) power of the bins in the $Z$ off-shell region in breaking the degeneracy in the sign of the coupling of the $Z$ to right-handed $b$ quarks.
When the $t\bar{t}$ background is not subtracted, the confidence contours shown in figure~\ref{fig:total} are limited by the theory error.
The three different error scenarios we consider are chosen as follows: 10\% roughly corresponds to the scale variation uncertainty at NLO, 5\% is a realistic projected improvement, and 2\% is only meant to illustrate how sensitive observable to theory and systematic uncertainties.

Figure~\ref{fig:subtracted_combined} shows the results for the charge asymmetry with the $t\bar{t}$ contribution subtracted using the different-flavor control region, see section~\ref{sec:subtracting_ttbar}. The left (right) panel of the figure has $\epchrg=65\%~(80\%)$.
Unlike the total charge asymmetry, the subtracted one is dominated by the statistical uncertainty and would, thus, benefit from larger integrated luminosities (see figure~\ref{fig:subtracted_lumi_projections} and its discussion).
Each panel in figure~\ref{fig:subtracted_combined} shows a comparison between the BDT cut discussed in sec.~\ref{sec:bdt} and the missing transverse energy cut.
The former does indeed improve the bound on the couplings albeit not dramatically. This is because the $\ttbar$ background only affects the subtracted asymmetry via the propagation of its uncertainty in the subtraction. Nevertheless, the improvement makes it clear that reducing the size of the $\ttbar$ background before subtracting it is important.

The individual contributions of the three $\mll$ regions below, around, and above the $Z$ pole are shown in figure~\ref{fig:subtracted_bins} for the same $\epchrg$ scenarios as in figure~\ref{fig:subtracted_combined} but only using the BDT cut.
Here again, this figure makes the role of the bins below the $Z$ pole in breaking the sign degeneracy on $g_{b,R}$ clear.
This region is where the $\gamma-Z$ interference is at its largest and thus has an enhanced sensitivity to the part of the squared amplitude with linear dependence on the $\zbb$ couplings.

Finally, the potential reach of the subtracted asymmetry which is limited by the statistical uncertainty is illustrated in figure~\ref{fig:subtracted_lumi_projections}.
The left panel of the figure uses the benchmark charge tagging efficiency $\epchrg=65\%$ while the right panel shows the improvement gained from an enhanced efficiency of $\epchrg=80\%$.
In both cases, we compare the baseline HL-LHC luminosity of $3~\abinv$ which was used in figures~\ref{fig:total}, \ref{fig:subtracted_combined}, and \ref{fig:subtracted_bins} to potential reach of the FCC-hh with $30\abinv$.
We stress again that we do not perform a dedicated FCC study. Rather, to get an approximate reach, we simply scale up the luminosity in our HL-LHC study with $\sqrt{s}=14$~TeV.
The solid red curves show the single-experiment reach for both the HL-LHC and FCC while the dashed grey curves show an approximate two-experiment combination (i.e. ATLAS + CMS).

\section{Summary and conclusions}
\label{sec:conclusion}
In this work, we propose a new channel at the LHC that is sensitive to the bottom quark charge asymmetry.
Previous efforts to directly probe the $Zb\bar{b}$ couplings at hadron colliders typically relied on reconstructing the $Z$ boson in the $b\bar{b}$ decay channel.
This method usually suffers from large QCD backgrounds; most notably from gluon splitting which is particularly acute when the $Z$ boson is boosted as is the case in LHCb.
Here, we propose that the $Z$ boson be reconstructed in the di-lepton channel in $Z$ production in association with a bottom quark pair.
This has the advantage of directly tagging the $b$-jets while at the same time reconstructing the $Z$ in a clean, leptonic, final state.
While the goal of achieving enough precision with the HL-LHC dataset of $3~\abinv$ to resolve the long-standing LEP anomaly is not borne out, the existing sign ambiguity of the right-handed coupling $g_{b,R}^Z$ can definitely be broken. Specifically, with a $b$-jet charge-tagging efficiency of 65\%~(80\%), the wrong sign is disfavored by $4\sigma~(6\sigma)$.
This can be achieved thanks to the ability of hadron colliders to probe the off-shell region as illustrated in figures~\ref{fig:total} and~\ref{fig:subtracted_bins}.

To mitigate the $\ttbar$ background which is by far the largest one, we compared the performance of missing transverse energy cut against that of a BDT classifier to reduce its contribution and enhance the sensitivity to the signal.
For the total charge asymmetry which includes a contribution from $\ttbar$, the theory error dominates over the statistical uncertainty as shown in figure~\ref{fig:total} where the left panel compares the bound using a theory error of $10\%$, roughly corresponding the seven-scale envelope at NLO to a realistic improvement of 5\% and an optimistic  one of 2\%.
However, since the $\ttbar$ pair also decays to different-flavor leptons, it can be completely subtracted using this control region.
In this case, the BDT cut is still useful as shown in figure~\ref{fig:subtracted_combined} since it reduces the induced statistical uncertainty on the subtracted asymmetry.
Furthermore, assuming that the same-flavor and different-flavor cross-sections have correlated theory and experimental uncertainties as discussed in section~\ref{sec:simulation}, we showed that the subtracted asymmetry is limited by the statistical uncertainty rather than by the theory error.
Accordingly, we explored various luminosity scenarios in figure~\ref{fig:subtracted_lumi_projections} which clearly illustrate the benefit of larger datasets especially if the optimistic charge-tagging efficiency of 80\% can be achieved.

Furthermore, combining our study with that of ref.~\cite{Gauld:2019doc} at LHCb which relies on the asymmetry in the rapidity difference of the jets initiated by a $b$ or $\bar{b}$ quark shown as a green band in figures~\ref{fig:total}, \ref{fig:subtracted_combined}, and \ref{fig:subtracted_lumi_projections} would further improve the reach of our study since the observables probe different directions in the $g_L$--$g_R$ coupling plane.
And, with the integrated luminosity target of the FCC-hh, the LEP anomaly could be disfavored with about 1$\sigma$ significance with our study alone. 
Numerically, setting $g_L$ to its SM value, a bound of about $g_{b,R}^Z\in [0.05,0.09]$ ([0.065,0.08]) can be achieved at the HL-LHC (extrapolated FCC-hh) dataset.

Future improvements especially from precision calculations to help reduce the overall theoretical uncertainty on the total asymmetry, and jet charge-tagging to increase cross-section in the case of the subtracted asymmetry would lead to much better bounds as illustrated in the right panels of figures~\ref{fig:subtracted_combined} and~\ref{fig:subtracted_lumi_projections}.
Finally, while we have ignored the contribution of charm-quark couplings in our current analysis their inclusion is necessary in a more detailed study. In particular, in a joint study of the bottom and charm charge asymmetry, one could take their ratios to reduce the theory error as was done in~\cite{Gauld:2019doc}. A joint study of the bottom and charm couplings is, moreover, motivated by the fact that tagging a bottom and charm flavored jets necessarily correlates observables involving both heavy flavors.
In any case, including the LHC observable we propose here in global fits would clearly improve the resulting bounds.

\acknowledgments

We would like to thank Pier Monni for collaborating in the early stages of this project.
We also thank Ayres Freitas, Christophe Grojean, Ulrich Haisch, Jasper Roosmale Nepveu, Emmanuel Stamou, and Doreen Wackeroth for useful discussions, Olivier Mattelaer for his help fixing an issue related to using the systematics module of \verb|MadGraph_aMC@NLO| in gridpack runs, and Ulrich Haisch and Christophe Grojean for their helpful comments on the manuscript.
Z.Q. would like to thank Zongguo Si and the group at Shandong university for helpful discussion and support during the work.
The work of Z.Q. was supported by the Helmholtz-OCPC fellowship program in the beginning, and later by the starting research fund of Hangzhou normal university. 
The Monte Carlo simulations were made possible thanks to the \texttt{theoc} cluster which is managed by the theory-group computing team at DESY. 
The work of F.B. was partially supported by the Deutsche Forschungsgemeinschaft (DFG, German Research Foundation) under grant 491245950 and under Germany’s Excellence Strategy — EXC 2121 ``Quantum Universe'' -- 390833306.
This work was performed in part at the Aspen Center for Physics, which is supported by National Science Foundation grant PHY-1607611.
This research was supported by the Munich Institute for Astro-, Particle and BioPhysics (MIAPbP) which is funded by the Deutsche Forschungsgemeinschaft (DFG, German Research Foundation) under Germany's Excellence Strategy – EXC-2094 -- 390783311.

\appendix

\section{Coupling dependence of the differential cross-section}
\label{app:dsigma}

The bulk of the contribution to the signal final state, $\bbll$, arises at $\mathcal{O}(\alpha_s^2\alpha^2)$ and thus contains only one electroweak boson propagator which mediates the lepton pair.
As such, the coupling dependence of the differential cross-section for this contribution is rather simple and follows from the sub-process discussed in section~\ref{sec:theory}.
The differential cross-section can be decomposed into three contributions that arise from $\gamma$--$\gamma$, $Z$--$Z$, and $\gamma$--$Z$ interference which we refer to as $d\sigma_\gamma$, $d\sigma_Z$, and $d\sigma_{\rm int}$, respectively. It can be further decomposed into symmetric and anti-symmetric pieces as discussed in section~\ref{sec:theory}.
The asymmetric piece vanishes, as expected, when integrating inclusively over the di-lepton phase-space.
It can be projected out by integrating against the sign of the scattering angle, $\theta$, defined in eqs.~\eqref{eq:asymmetric_observable} and ~\eqref{eq:cs_angle_ll_frame}.
The three components with explicit dependence on the couplings of the leptons and $b$-quarks to the electroweak bosons for the symmetric contribution are,
\begin{equation}
\begin{split}
\frac{d\sigma_\gamma}{dm_{\ell\ell}} & = F (m_{\ell\ell}) \frac{4\,Q_\ell^2\,Q_b^2}{\propg ^2}\\
\frac{d\sigma_Z}{dm_{\ell\ell}} & = F (m_{\ell\ell})\frac{1}{(\sin^2\theta_{W}\cos^2\theta_{W})^2}\frac{(g_{b,L}^2 + g_{b,R}^2)(g_{e,L}^2 +g_{e,R}^2) }{\propz^2}\\
\frac{d\sigma_{\rm int}}{dm_{\ell\ell}} & = F (m_{\ell\ell}) \frac{2\,Q_\ell\,Q_b}{\sin^2\theta_{W}\cos^2\theta_{W}}\frac{(g_{b,L} + g_{b,R})(g_{e,L} +g_{e,R}) }{\propz \propg}.
\end{split}
\label{eq:dsigma_s}
\end{equation}
Here, $\propg$ and $\propz$ are the photon and $Z$ propagator denominators and are defined in the main text (see section~\ref{sec:theory}), and $\theta_W$ is the weak mixing angle.
Likewise, with a common factor $H(m_{\ell\ell})$, the asymmetric contribution reads,
\begin{equation}
\begin{split}
\frac{d\sigma^{A}_{\gamma}}{dm_{\ell\ell}} & = 0\\
\frac{d\sigma^{A}_Z}{dm_{\ell\ell}} & = H (m_{\ell\ell})   \frac{1 }{(\sin\theta_{W}^2\cos\theta_{W}^2)^2}\frac{(g_{b,L}^2 - g_{b,R}^2)(g_{e,L}^2 -g_{e,R}^2) }{\propz ^2}\\
\frac{d\sigma^{A}_{\rm int}}{dm_{\ell\ell}} & = H (m_{\ell\ell}) \frac{2\,Q_\ell\,Q_b }{\sin\theta_{W}^2\cos\theta_{W}^2}\frac{(g_{b,L} - g_{b,R})(g_{e,L} -g_{e,R}) }{\propz \propg}.
\end{split}
\label{eq:dsigma_a}
\end{equation}
The asymmetric cross section is measured through any $b$ and $\bar b$ asymmetric observable.
The symmetric and asymmetry differential cross section for signal $gg\to b\bar b\ell^-\ell^+$ process reads,
\begin{equation}
A(m_{\ell\ell})=\frac{d\sigma^A_{tot}}{d\sigma_{tot}} = \frac{d\sigma_\gamma^A + d\sigma^{A}_Z +  d\sigma^{A}_{\rm int} } {d\sigma_\gamma + d\sigma_Z + d\sigma_{\rm int}}.
\label{eq:asymexp}
\end{equation}

\begin{figure}[t]
	\centering
	\begin{minipage}{.45\textwidth}
		\centering
		\includegraphics[width=0.98\linewidth]{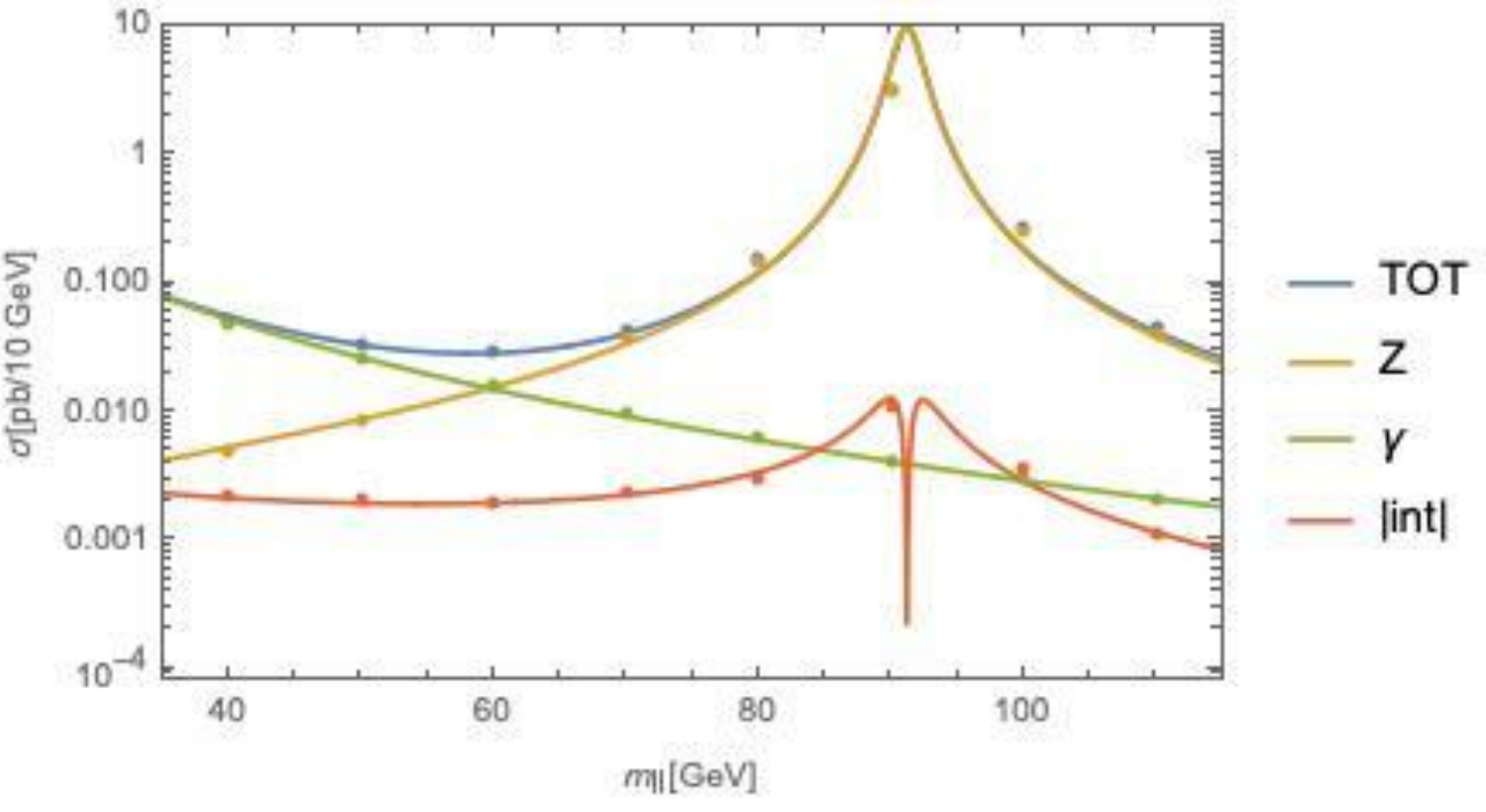}
	\end{minipage}%
	\begin{minipage}{.45\textwidth}
		\centering
		\includegraphics[width=0.98\linewidth]{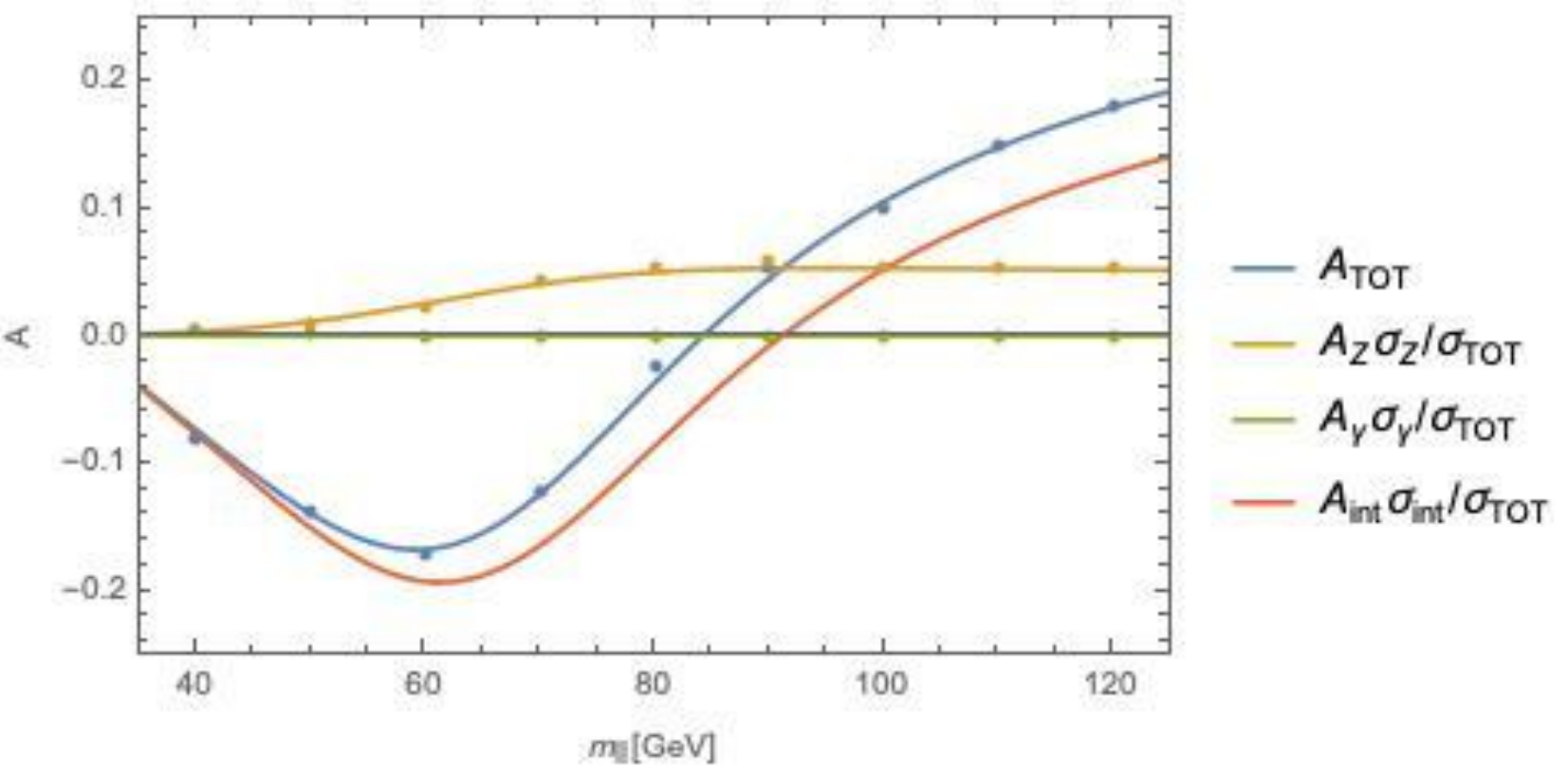}
	\end{minipage}%
	\caption{The left (right) panel shows the differential cross-section (asymmetry) for each of the three components that arise from $\gamma$--$\gamma$, $Z$--$Z$, and $\gamma$--$Z$ interference along with their sum. These are shown as solid curves. The dots show the parton-level cross-sections computed with \texttt{MadGraph}, see text for details.}
	\label{fig:ggbbll_mll}
\end{figure}
The differential cross-section and asymmetry with respect to the di-lepton invariant mass are shown in figure~\ref{fig:ggbbll_mll}.
The solid curves are obtained from the semi-analytic form in eqs.~\eqref{eq:dsigma_s} and~\eqref{eq:dsigma_a} after extracting the common factors $F(\mll)$ and $H(\mll)$ from the \texttt{MadGraph} simulation.
The differential cross-section and asymmetry from this simulation are shown by dots in the figure.
One can see that the $Z$-mediated contribution is dominant above $m_{\ell\ell}\gtrsim 60$ GeV and that below this invariant mass, the photon-mediated contribution dominates.
Thus, the interference between the photon and $Z$ mediated contributions is only relevant around this cross-over point.
Calculation as well as simulation show that the signal asymmetry is zero at an invariant mass $m_{\ell\ell}\sim 84$ GeV given SM couplings. This is due to a cancellation between the terms that linear and quadratic in the couplings. Interestingly, this value holds even after the full simulation including detector effects; see the right panel in figure \ref{fig:Ameasure}.

\section{Numerical fits and results from full simulation}
\label{app:glgrfit}
The results of our fit for the coefficients of eq.~\eqref{eq:msquared_comps} are shown in table~\ref{tab:numfit}.
\begin{table}[h]\centering\renewcommand{\arraystretch}{1.2}
\begin{tabular}{c|ccccc}\toprule[1pt]
& const. & $g_L + g_R$ & $g_L^2 + g_R^2$ & $g_L - g_R$ & $g_L^2-g_R^2$\\\cline{2-6}
Bin [GeV] & $F_0$ & $F_1$ & $F_2$ & $H_1$ & $H_2$\\
\midrule
35 -- 45 & 9.524 & 0.9347 & 7.915 & 0.9898 & 0.2415\\
45 -- 55 & 6.248 & 0.9759 & 15.90 & 1.346 & 0.6219\\
55 -- 65 & 4.527 & 1.145 & 31.15 & 1.730 & 0.9969\\
65 -- 75 & 4.096 & 1.483 & 70.17 & 2.334 & 2.926\\
75 -- 85 & 8.066 & 2.014 & 282.8 & 4.328 & 12.20\\
85 -- 95 & 92.84 & 1.501 & 3907 & 0.5029 & 180.9\\
95 -- 105 & 8.723 & -2.619 & 306.8 & -4.808 & 15.81\\
105 -- 115 & 2.249 & -0.8266 & 52.47 & -1.724 & 2.650\\
115 -- 125 & 1.235 & -0.4347 & 21.00 & -1.004 & 1.133\\
\bottomrule[1pt]
\end{tabular}
    \caption{Bin-by-bin fit results for the coefficients of eq.~\eqref{eq:dsigma_dmll_coeffs} after all selection cuts defined in eq.~\eqref{eq:final_acceptance_cuts} are applied but without any flavor or charge tagging efficiencies. These are applied at a later stage in the analysis. }
\label{tab:numfit}
\end{table}

\begin{figure}[b]
    \centering
    \includegraphics[width=0.45\linewidth]{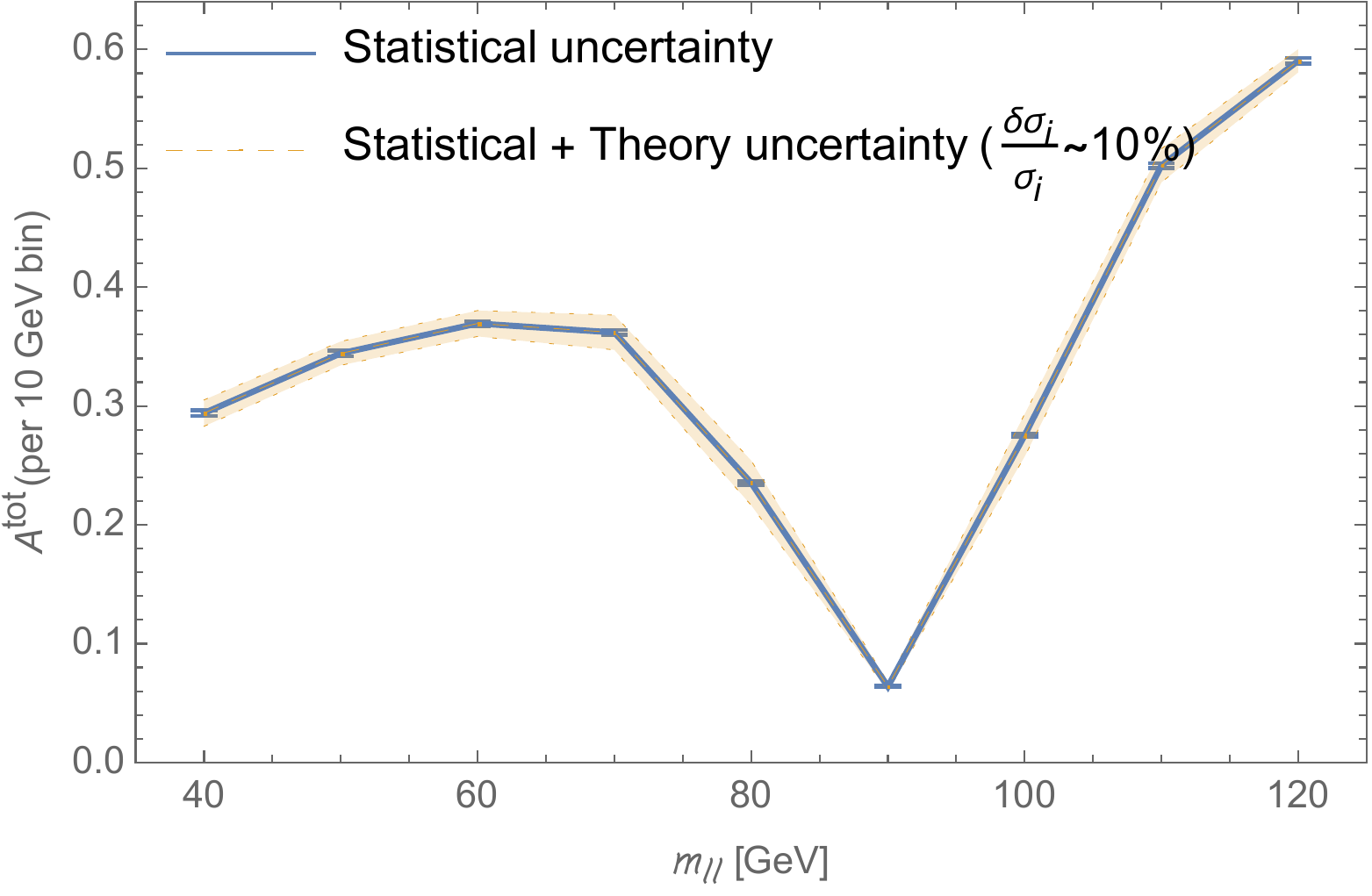}
    \includegraphics[width=0.45\linewidth]{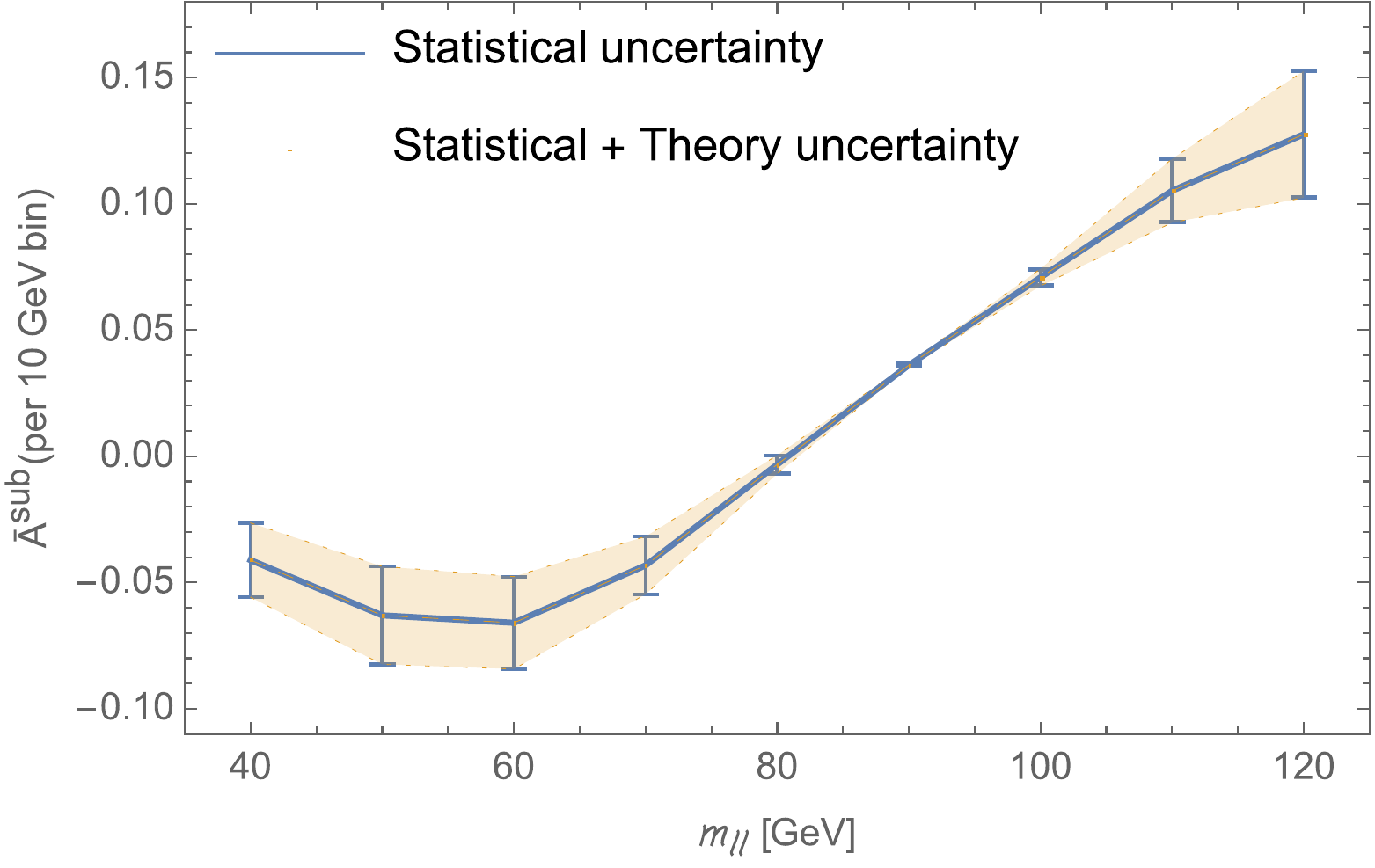}
    \caption{The left (right) panels show the total (subtracted) asymmetry for $\ttbar$ ($b\bar{b}\ell^+\ell^-$). For the total asymmetry, we use $\delta\sigma/\sigma=10\%$ as a rough NLO estimate on the $\ttbar$ differential cross-section. The error bars show the statistical uncertainty with $3~\abinv$ and the cream bands show the combined statistical and LO theory errors; see sec.~\ref{sec:theory_uncertainty} and table~\ref{tab:sm_tt_and_bbz}.}
    \label{fig:Ameasure}
\end{figure}

Table~\ref{tab:sm_tt_and_bbz} shows the SM cross-sections and asymmetries for the signal and the $\ttbar$ background after the selection cuts of eq.~\eqref{eq:final_acceptance_cuts} are applied but without any flavor or charge tagging efficiencies. In addition, the relative theory error on the cross-sections and asymmetries at LO are also shown.
\begin{table}[]\centering\renewcommand{\arraystretch}{1.2}
	\begin{tabular}{c | c c c c | c c c c }\toprule[1pt]
		&\multicolumn{4}{c|}{$\bbbarz$}&\multicolumn{4}{c}{$\ttbar$}\\
		Bin [GeV] & $\sigma$ [fb] & $\dfrac{\delta\sigma}{\sigma}$ [\%] & $A$ & $\dfrac{\delta A}{A}$ [\%] &  $\sigma$ [fb] & $\dfrac{\delta\sigma}{\sigma}$ [\%] & $A$ & $\dfrac{\delta A}{A}$ [\%] \\\midrule
		35.0 -- 45.0 & 10.7 & 28 & -0.0371 & 0.86 & 35.8 & 21 & 0.393 & 1.5\\
		45.0 -- 55.0 & 8.85 & 27 & -0.0613 & 0.98 & 42.4 & 21 & 0.428 & 1.6\\
		55.0 -- 65.0 & 10.0 & 26 & -0.0579 & 1.1 & 45.7 & 21 & 0.461 & 1.8\\
		65.0 -- 75.0 & 16.7 & 25 & -0.0404 & 1.4 & 46.5 & 21 & 0.499 & 1.7\\
		75.0 -- 85.0 & 60.6 & 24 & -0.00338 & 1.6 & 45.3 & 21 & 0.534 & 1.7\\
		85.0 -- 95.0 & 828. & 23 & 0.0369 & 1.5 & 42.7 & 21 & 0.568 & 1.7\\
		95.0 -- 105.0 & 67.3 & 23 & 0.0726 & 1.8 & 39.1 & 21 & 0.604 & 1.6\\
		105.0 -- 115.0 & 12.4 & 22 & 0.104 & 2.0 & 35.2 & 21 & 0.634 & 1.5\\
		115.0 -- 125.0 & 5.35 & 22 & 0.128 & 2.0 & 29.7 & 21 & 0.669 & 1.4\\
		\bottomrule[1pt]
	\end{tabular}
\caption{The LO cross-section and asymmetries along with the LO theory error for the signal, $b\bar{b}\ell^+\ell^-$, and the $\ttbar$ background.}
\label{tab:sm_tt_and_bbz}
\end{table}
The total and subtracted asymmetries are plotted in figure~\ref{fig:Ameasure} where the statistical error with the HL-LHC dataset of $3~\abinv$ is shown with with vertical error bars and the combined statistical and theory error is shown as a cream-colored band. For the total asymmetry, the theory error is given in eq.~\eqref{eq:theory_uncertainty_on_total_asymmetry} and in the figure, we use $\delta\sigma/\sigma=10\%$ as a rough estimate of the NLO error on the differential cross-section.

\section{Dependence of the LEP and LHCb bounds on the couplings}
\label{sec:comparison_bounds}

At LEP and $e^+e^-$ colliders in general, the dependence of the two observables of interest discussed above, in terms of the left- and right-handed $Zb\bar b$ couplings, $g_{b,L}$ and $g_{b,R}$, is given by,
\begin{equation}
\begin{split}
R^0_{b}=R^{\rm SM}_b + \delta R_b,\qquad\delta R_b & \sim  g^{\rm SM}_{b,L} \delta g_{b,L} + g^{\rm SM}_{b,R} \delta g_{b,R},\\
A_{FB} = A^{\rm SM}_{FB}+ \delta A_{FB},~~\delta A_{FB} & \sim
g^{\rm SM}_{b,R}\delta g_{b,L}- g^{\rm SM}_{b,L}\delta g_{b,R}\,,\\
\end{split}
\label{eq:thm_full}
\end{equation}
where we only keep the leading terms in the small deviations from the SM couplings, $\delta g_{b,L/R}$, and drop overall constant factors.
Equation~\eqref{eq:thm_full} nicely illustrates that these two observables lead to orthogonal constraints in the $g_L-g_R$ plane near the SM point.
At hadron colliders, however, the inclusive rate is dominated by QCD backgrounds by at least an order of magnitude, even around the $Z$ pole.
This leads to worse sensitivity and destroys the orthogonality to $\delta R_b$ in eq.~\eqref{eq:thm_full}. Expanding to linear order in the coupling deviations around the SM point and in the small ratio $\alpha^2/\alpha_s^2$ (see ref.~\cite{Gauld:2015qha} for the unexpanded expression), the change in the forward-backward and forward-central asymmetries at hadron colliders is given by,
\begin{equation}
\begin{split}
\delta A_{F[B/C]} & \sim \frac{\alpha^2}{\alpha_s^2}\left(g_{b,L}^\textrm{SM} \delta g_{b,L} - g_{b,R}^\textrm{SM} \delta g_{b,R}\right)\,.
\label{eq:thm_delta}
\end{split}
\end{equation}

\bibliographystyle{JHEP.bst}
\bibliography{references}

\end{document}